\documentclass[journal = jpccck, manuscript = article]{achemso}
\setkeys{acs}{usetitle = true}
\usepackage{amsmath}
\usepackage{times}
\usepackage{xcolor}
\usepackage{booktabs}
\usepackage{mathptmx}
\usepackage[version=4]{mhchem}
\usepackage{lineno}
\usepackage{gensymb}
\usepackage{multirow}
\usepackage{dcolumn}
\usepackage{wrapfig}
\usepackage{xr}

\makeatletter
\newcommand*{\addFileDependency}[1]{% argument=file name and extension
  \typeout{(#1)}
  \@addtofilelist{#1}
  \IfFileExists{#1}{}{\typeout{No file #1.}}
}
\makeatother

\newcommand*{\myexternaldocument}[1]{%
    \externaldocument{#1}%
    \addFileDependency{#1.tex}%
    \addFileDependency{#1.aux}%
}
\myexternaldocument{ZThiolPEG}

\newcommand{\angstrom}{\mbox{\normalfont\AA}}

\title{Heat Transfer in Gold Interfaces Capped with Thiolated Polyethylene Glycol: A Molecular Dynamics Study}
\author{Sydney A. Shavalier}
\affiliation[University of Notre Dame]{251 Nieuwland Science Hall, Department of Chemistry and Biochemistry, \\
University of Notre Dame, Notre Dame, Indiana 46556}
\author{J. Daniel Gezelter}
\email{gezelter@nd.edu}
\affiliation[University of Notre Dame]{251 Nieuwland Science Hall, Department of Chemistry and Biochemistry, \\
University of Notre Dame, Notre Dame, Indiana 46556}

\begin{document}

\begin{tocentry}
\center\includegraphics[width=7.4cm]{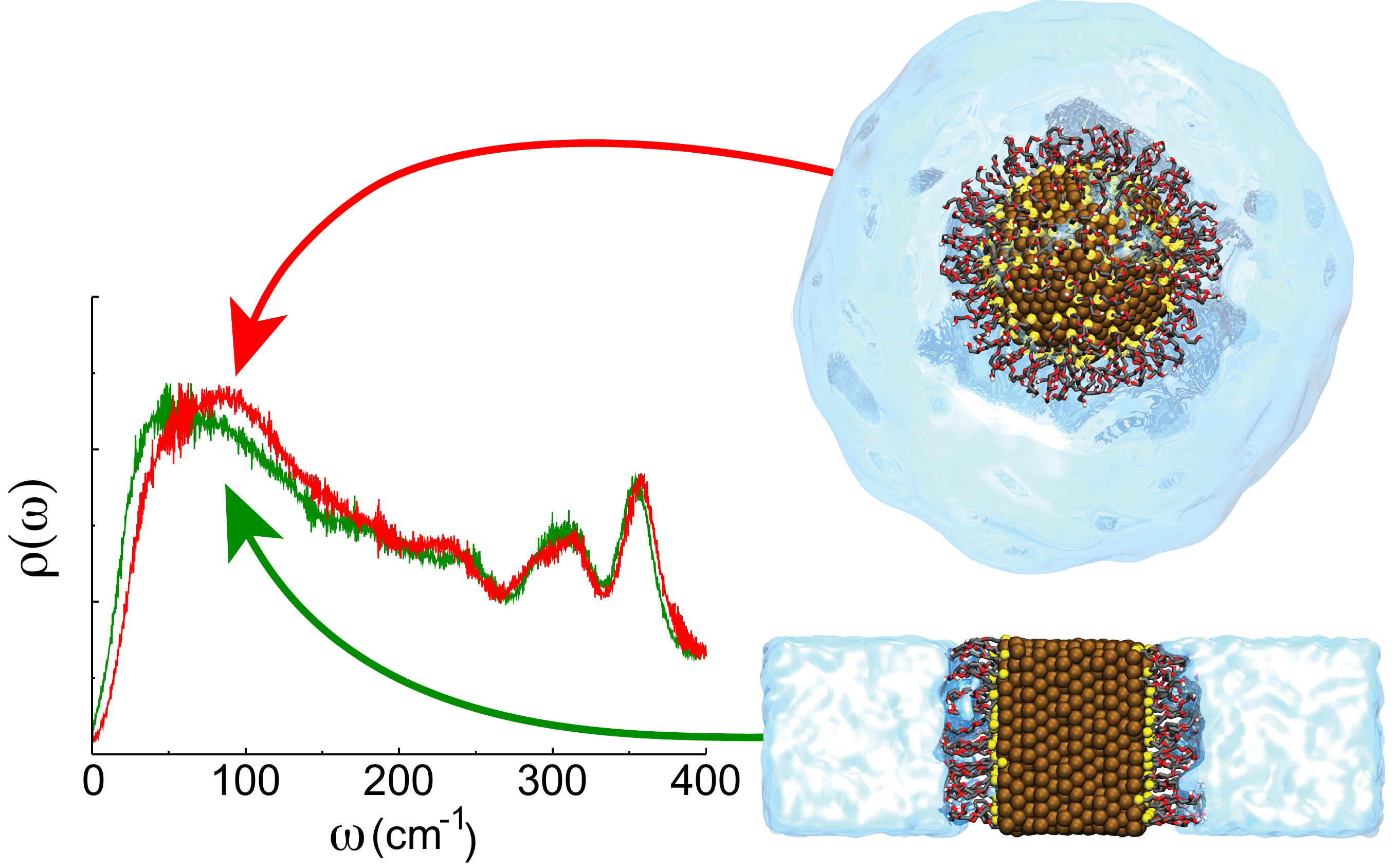} 
\end{tocentry}

\begin{abstract}
Reverse non-equilibrium molecular dynamics (RNEMD) simulations were used to study heat transport in solvated gold interfaces which have been functionalized with a low molecular weight thiolated polyethylene glycol (PEG). The gold interfaces studied included (111), (110), and (100) facets, as well as spherical nanoparticles with radii of 10 and 20 \AA{}. The embedded atom model (EAM) and the polarizable density-readjusted embedded atom model (DR-EAM) were implemented to determine the effect of metal polarizability on heat transport properties. We find that the interfacial thermal conductance values for thiolated PEG capped interfaces are higher than  those for pristine gold interfaces.  Hydrogen bonding between the thiolated PEG and solvent differs between planar facets and the nanospheres, suggesting one mechanism for enhanced transfer of energy, while the covalent gold sulfur bond appears to create the largest barrier to thermal conduction. Through analysis of vibrational power spectra, we find an enhanced population at low frequency heat carrying modes for the nanospheres, which may also explain the higher mean $G$ value.

\end{abstract}

\section{Introduction}

 The unique optical properties of gold nanostructures make them useful candidates for biomedical applications such as photothermal therapies.\cite{jbv19, xh10, xb20, wy19, ks19, zjl11} The localized surface plasmon resonance allows excitation of conduction band electrons upon irradiation with light at the same frequency as the plasmon resonance, giving gold nanostructures the ability to absorb light at particular wavelengths in the near IR.

Due to their similarities to sulfur-containing biological molecules, thiols and thiolates that functionalize gold nanoparticles have been studied extensively.\cite{lbp15, cam96, xz12, cp20, nrss21, jl07} The thiol functional group is present in the amino acid cysteine, and cysteine adsorption onto gold has been studied both computationally~\cite{rdf04, gy18, sm16} and experimentally.\cite{zpl06, gy18} Upon reaction with colloidal gold, thiols have been shown to bind to the gold via a strong Au-S bond, releasing hydrogen gas by breaking the S-H bond. In one study, the presence of hydrogen gas was confirmed using gas chromatography.\cite{jem12} Alternatively, thiolate-capped gold nanoparticles can be synthesized via a ligand exchange reaction~\cite{ghw05} or a two phase reduction.\cite{mb94}
For this work, we are concentrating on a low molecular weight thiolated poly(ethylene glycol) (PEG) as a ligand. PEG is a biocompatible, water-soluble polymer with a molecular length that can be tailored for a variety of uses, notably as a bioconjugation agent.\cite{sz95}  The transfer of heat through nanomaterials functionalized with thiolated PEG is important in potential application to both photothermal and gene therapies.\cite{zk14,lf13,lc18} 

Significant experimental~\cite{hh13} and simulation studies\cite{ks13, ks16, sk11, smn20, ss20} have contributed to our current understanding of heat transfer through thiolate-capped gold interfaces. Using time-domain thermoreflectance, Harikrishna \textit{et al.} determined the interfacial thermal conductance ($G$) from a planar gold film capped with a variety of alkanethiol self-assembled monolayers (SAMs) that were solvated in water. They concluded that varying the terminal group of the alkanethiols changed the conductance, but the change could also be caused by the ability of water molecules to penetrate between the alkanethiol chains.\cite{hh13} Their measured values of $G$ were supported by molecular dynamics (MD) simulations of the same systems by Hung \textit{et al.}\cite{sh16} They found that the crowding of water molecules between the alkanethiols did indeed influence $G$, suggesting that it could be tuned by controlling, either via the terminal functional group or the SAM binding density, the ability of water to aggregate near the terminal end of the ligands.\cite{sh16} 

There is significant debate over how the size of a solvated gold nanosphere influences thermal transport due to changes in surface curvature. Tascini \textit{et al.}\cite{ast17} and Jiang \textit{et al.}\cite{mj22} determined that in systems of Lennard-Jones spherical nanoparticles solvated in a Lennard-Jones fluid, $G$ increased with decreasing particle size. The prevailing argument for this trend is that the increase in undercoordinated surface atoms in smaller nanospheres creates a larger angle of contact between the surface atoms and the surrounding solvent. However, a study of gold nanospheres solvated in hexane concluded that while $G$ increased from a flat Au(111) surface to spherical particles, there was very little size dependence over a range of particle radii from 20 \AA{} to 40 \AA{}.\cite{kms14} Additionally, a study of gold nanoparticles capped with various alkanethiolates and solvated in hexane found that ligand structure and rigidity had a larger effect on the value of $G$ than the size of the particle core.\cite{ks16}  In studies of pristine spherical gold nanoparticles solvated in water, the increase in conductance due to gold surface curvature was explained by the increase in the number of water molecules (energy carriers) close to the gold surface.\cite{baw22, lep23, og22} These results suggest that all three effects, such as coordination number of surface gold atoms, metal-to-ligand coupling, and metal-to-solvent interactions may impact interfacial thermal conductance. 

Simulations of heat conduction through flat Au(111)-hexane interfaces with mixed-chain alkanethiolate ligands showed that ligand chain alignment has a significant effect on interfacial thermal conductance. Solvent molecule entrapment, in which hexane molecules were embedded between adjacent alkanethiolate ligands, substantially affected $G$.  Orientational alignment between the solvent and ligand was found to increase thermal transfer from  the ligand to the entrapped solvent molecules.\cite{ks13} Stocker and Gezelter and Harikrishna \textit{et al.} separately determined via simulation\cite{ks13} and experiment\cite{hh13} that interfacial thermal conductance is not directly dependent on the ligand chain length in systems containing thiolate-capped gold.  This can potentially be explained by ballistic heat transport through the ligand chains on the metal surface.\cite{hdp16}

Few previous simulation studies have considered the role that metal polarizability plays in these thermal transport processes. In this work, we utilize the density-readjusted embedded atom model (DR-EAM) from Bhattarai \textit{et al.}\cite{hb19}, which is a polarizable metal potential. The metal charges are propagated along with the atomic positions and velocities in simulations where DR-EAM is implemented. Two previous studies\cite{hb20, ss22} found that $G$ increases when DR-EAM is implemented in systems of  pristine  Au(111) solvated in water. %However, the situation with bound ligands is less clear. We previously determined that with the addition of 0.12 M sodium citrate, metal polarizability had little to no effect on $G$ in planar systems. We also found that the influence of metal polarizability on the interfacial thermal conductance of citrate-capped gold interfaces varies depending on the morphology of the gold.\cite{ss22}

In this work, we have simulated thermal transport through planar Au surfaces and small Au nanospheres which have been capped with a fixed surface density of a low molecular weight thiolated PEG, \ce{S(-CH2CH2O-)_3H}. To study the effect of different surface facets, we used planar gold systems that expose the (100), (111), and (110) facets of crystalline gold to the ligand and solvent. Gold nanospheres ($r = 10$ and $r = 20$ \AA{}) were simulated with the same surface ligand density, but they were encased in a solvent cloud rather than a periodic box. Details on the systems and the methods for generating thermal transport are presented in the following sections.

\section{Methods}

 To simulate thermal transport, we utilized velocity shearing and scaling reverse nonequilibrium MD (VSS-RNEMD).\cite{sk12} In this method, an unphysical kinetic energy (or momentum) flux is applied by scaling (or shearing) the particle velocities in two separated regions of the simulation. These regions are typically rectangular slabs in planar geometries and concentric spheres in the nonperiodic geometries.\cite{kms14} Temperature (or velocity) gradients form between the regions in response to the applied fluxes. To generate the fluxes, velocities are modified for particles $i$ in the cold slab and particles $j$ in the hot slab:
\begin{eqnarray}
    \mathbf{v}_i & \leftarrow c \cdot (\mathbf{v}_i-\langle \mathbf{v}_c\rangle ) + (\langle \mathbf{v}_c\rangle +  \mathbf{a}_c) \\
    \mathbf{v}_j & \leftarrow h \cdot (\mathbf{v}_j-\langle \mathbf{v}_h\rangle ) + (\langle \mathbf{v}_h\rangle + \mathbf{a}_h)
\end{eqnarray}
where $\langle \mathbf{v}_h\rangle$ and $\langle \mathbf{v}_c\rangle$ describe the instantaneous center of mass velocities for all molecules occupying the hot and cold slabs, respectively. At every time step, velocities are scaled by two variables, $h$ and $c$, while $\mathbf{a}_c$ and $\mathbf{a}_h$ are shearing terms that alter the relative mean velocities between the slabs. The applied flux values and conservation of energy and momentum  provide simple solutions for the scaling ($c$ and $h$) and shearing ($\mathbf{a}_c$ and $\mathbf{a}_h$) variables.  

In standard NEMD methods, a gradient is applied using thermostats in two different regions, and a flux is calculated from the changes made to the velocities. In RNEMD methods, the flux is applied and the system's response (either a thermal gradient or temperature drop at an interface) is measured. In practice, either approach can be used with Fourier's law to compute thermal transport properties.  However, applied flux methods like the one used here can be made to satisfy conservation of energy and linear momentum, and can be added to any molecular dynamics integration method.

The effects of gold morphology and surface polarizability on the solvent thermal conductivity and interfacial thermal conductance were investigated. We determined the organization of the ligand layer and its contact with both the metal and solvent, as well as the hydrogen bonding density between the ligand and solvent, the magnitude of metal-to-ligand and ligand-to-solvent coupling via Bhattacharyya coefficient (BC) calculations, and the vibrational power spectra.

\subsection{Force Fields}

For metallic interactions, we used the embedded atom model (EAM) parametrized by Zhou \textit{et al.}\cite{xwz04} for nonpolarizable systems and the density-readjusting embedded atom model (DR-EAM) by Bhattarai \textit{et al.}\cite{hb19} for polarizable systems. These two models give nearly identical bulk metallic properties for gold, but differ in how they interact with charged and polar species  at the surface of the metal.
All systems were solvated in liquid water which was simulated using the SPC/E water model.\cite{hjcb87} Water molecules were simulated as rigid bodies; therefore, no bond or angle constraints were imposed.

The thiolated PEG parameters were adapted from multiple sources. The TraPPE united-atom force field was utilized for most Lennard-Jones parameters, bond lengths, and bend and torsion parameters.\cite{nl05, bc01, js04}  We implemented a fully flexible thiolated PEG model, and bond stretch force constants were adapted from OPLS-UA and OPLS-AA.\cite{sjw86, wlj96} The cross interactions between gold and other atom types were adapted from Schapotschnikow \textit{et al.}\cite{ps07}

Interactions between gold and water were modified from the Mie potential from Dou \textit{et al.}\cite{yd01}  The details of all force field parameters are provided in the Supporting Information.

\subsection{Simulation Protocol}
The following systems of interest were constructed: planar interfaces of single-crystalline gold exposing the (111), (110), and (100) facets as well as spherical gold nanoparticles (10 \AA{} and 20 \AA{} radii).  These five systems were studied using both nonpolarizable (EAM) and polarizable (DR-EAM) metallic models, yielding  a total of ten unique systems.  All metallic systems were constructed with a lattice constant of 4.08 \AA{}. For each system, four statistically independent samples were created by resampling the positions of the thiolated PEG molecules at the gold surfaces using Packmol.\cite{pack}  In all systems, the sulfur atoms of the thiolated PEG molecules were initially placed within 3 \AA{} of the gold surface. Orientational preferences were enforced in the initial configurations by placing the terminal hydroxyl hydrogen at least 9 \AA{} from the same surface.  On nanoparticles (centered at the origin), the sulfur atoms were placed inside a sphere that was 1-3 \AA{} larger than the radius of the nanosphere, while orientations were enforced by placing hydroxyl hydrogens at least 10 \AA{} farther away from the nanoparticle surface. These yielded relatively uniform initial orientations for all of the ligand molecules.  Built-in OpenMD utilities were used to combine the metal and ligand components and to solvate the systems. Each system then underwent steepest descent structural optimization prior to equilibration.

Initial atomic velocities were sampled from a Maxwell-Boltzmann distribution at 300~K. Planar systems then underwent structural relaxation for 20 ps in the canonical (NVT) ensemble, followed by a 400 ps equilibration in an isobaric-isothermal (NPT) ensemble with  characteristic time constants, $\tau_{\text{barostat}} = \text{5000 fs}$ and $\tau_{\text{thermostat}} = \text{100 fs}$  to adjust the volume of the simulation cell. This was followed by 200 ps of thermal equilibration (NVT) with $\tau_{\text{thermostat}} = \text{100 fs}$ and 200 ps of equilibration in the microcanonical (NVE) ensemble. These systems were subsequently simulated for 100 ps in the NPAT (constant normal pressure with the surface area of the interface held constant) ensemble.  A Nos\'e-Hoover thermostat\cite{wgh85} and modified Nos\'e-Hoover-Andersen\cite{sm93} barostat were used for constant temperature and pressure ensembles, respectively. A time step of 1 fs was used for all simulations.

Spherical gold nanoparticles were constructed by carving out a spherical region from an fcc lattice ($a = 4.08$ \AA{}) with a radius of either 10 or 20 \AA{}. The atoms were assigned velocities from a Maxwell-Boltzmann distribution at 5~K, and the  pristine  nanoparticles were incrementally equilibrated at 5, 100, 200~K, and and last 300~K in a Langevin Hull,\cite{cfv11} which can sample constant temperature and pressure ensembles in nonperiodic systems. As described above, Packmol and OpenMD were used to resample ligand positions and combine components.\cite{pack,openmd} One gold atom closest to the origin in each nanoparticle was given an artificially large mass to prevent the nanoparticle from drifting. These nonperiodic systems were first equilibrated for 200 ps using the Langevin Hull integrator\cite{cfv11} and a resampling of velocities from a Maxwell-Boltzmann distribution at 300~K every ps. This was followed by an additional 250 ps of equilibration using the Langevin Hull integrator without velocity resampling.

\begin{figure}[h]
    \centering
    \includegraphics[width=\linewidth]{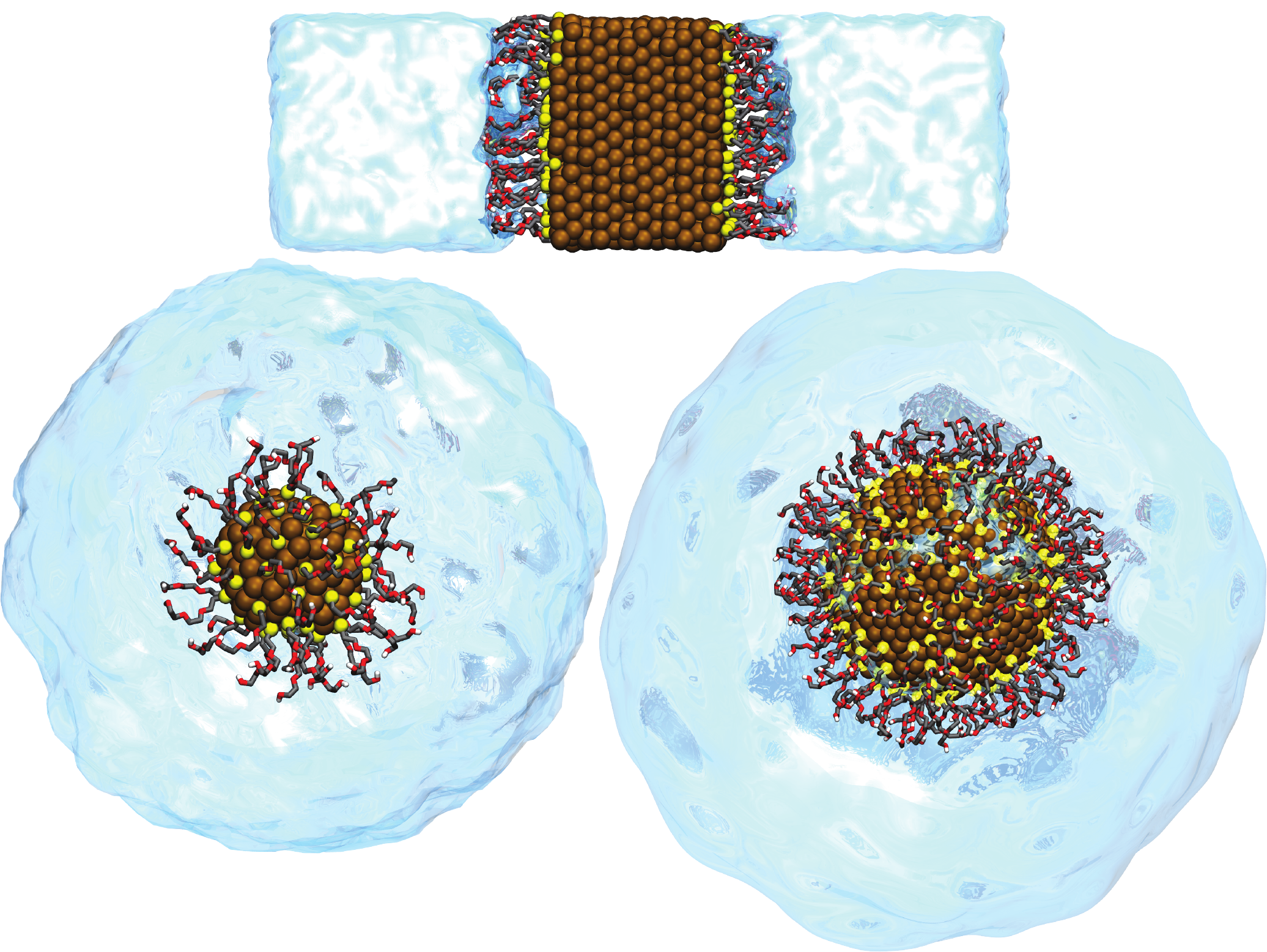}
    \caption{Representative configurations of the simulated systems. Top: Au(111) facet functionalized with thiolated PEG and solvated in SPC/E water. Bottom left: A gold nanosphere with a 10 \AA{} radius. Bottom right: A gold nanosphere with a 20 \AA{} radius. Gold and sulfur atoms have been represented with van der Waals spheres. The rest of the united atom thiolated PEG ligands are shown with bonds. The SPC/E water occupies the remaining volume. Detailed specifications of these systems are provided in the Supporting Information.
    \label{fig:systems}} 
\end{figure}

All systems were built with similar thiolated PEG surface densities (4.70 molecules/nm$^2$ for spherical systems with a gold radius of 10 \AA{}, 4.66 molecules/nm$^2$ for spherical systems with a gold radius of 20 \AA{}, 4.81 molecules/nm$^2$ for planar (111) facets, 4.67 molecules/nm$^2$ for planar (110) facets, and 4.63 molecules/nm$^2$ for planar (100) facets. These densities amount to an approximate coverage of 1 thiolated PEG molecule for every 3 surface gold atoms. Representative configurations of the systems are listed in Figure \ref{fig:systems}.

\clearpage

After equilibration, a thermal flux was applied to all systems using the velocity shearing and scaling (VSS) RNEMD methodology developed by Kuang and Gezelter.\cite{sk12}. The magnitude of the thermal flux was chosen so that the heat rate through the metal/solvent interface was approximately equal in all systems. The applied thermal flux  was $8.8 \times 10^{-6}$ $\mathrm{~kcal~mol^{-1} \angstrom^{-2}~fs^{-1}}$ (6100 $\mathrm{~MWm^{-2}}$)  for nanoparticles with a radius of 10 \AA{}, $3.6 \times 10^{-6}$ $\mathrm{~kcal~mol^{-1} \angstrom^{-2}~fs^{-1}}$ (2500 $\mathrm{~MWm^{-2}}$)  for the 20-\AA{} nanoparticles, and $5.0 \times 10^{-6}$ $\mathrm{~kcal~mol^{-1} \angstrom^{-2}~fs^{-1}}$ (3500 $\mathrm{~MWm^{-2}}$)  for all planar systems. The thermal flux was applied for 5 ns in all systems, but for the nanospheres, only the last ns was used for data collection.

\subsection{Calculation of Thermal Transport Properties}
To calculate the thermal conductivity $(\lambda)$ in the bulk solvent phase, we can use Fourier's law
\begin{equation}
\mathbf{J} = -\lambda \nabla T ,
\label{eq:fourier}
\end{equation}
where $\mathbf{J}$ is the thermal flux and $\nabla T$ is the temperature gradient. In RNEMD simulations, $\mathbf{J}$ is an unphysical applied perturbation, and the system responds by creating a thermal gradient. In planar systems, the thermal flux is applied in a direction normal to the interface (in this case the $z$-axis). From the applied flux, $J_z$, we can also calculate the interfacial thermal conductance
\begin{center}
\begin{equation}
    G=\left(\frac{J_z}{\Delta T}\right)
    \label{eq:periodicg}
\end{equation}
\end{center}
where $\Delta T$ is the temperature difference calculated  across the interface.

In planar systems, thermal gradients are computed using histograms containing local information about the temperature in 1.8 \AA{} wide bins which span the width and breadth of the system. We define the interfacial region, i.e. the region spanning the metal / ligand / solvent interface, from the last bin which contains only gold atoms to the first bin which contains only water. In computing the interfacial thermal conductance, the temperature jump, $\Delta T$, is measured across the entire interfacial region.  Calculations of solution-phase thermal conductivity $(\lambda)$ include only bins in the pure solvent phase. 

For spherical nanoparticles, we use a series approximation for the Kapitza resistance for concentric spherical shells
\begin{equation}
R_K = \frac{1}{G} = \frac{1}{q_r} \sum_{i} 4 \pi r_i^2 \left( T(r_{i+1}) - T(r_i) \right)
    \label{eq:nonperiodicg}
\end{equation}
The idea here is that heat must pass through successive spherical shells with an identical radial heat rate, $q_r$, rather than the applied radial flux, $J_r$. We accumulate statistics on 1.25 \AA{} wide shells, but here each shell has a different surface area and volume. Shell $i$ has an inner radius $r_i$, and the average temperature of that shell is $T(r_i)$. The interfacial region is defined as a finite-width region starting at a shell containing only gold atoms and ending at a shell containing only water. In a previous study, we showed that varying the width of the interface used for calculating $G$ remains relatively stable as long as it spans the entire interfacial region.\cite{ss22}
Transfer of heat from one shell to another contributes a small amount to the Kapitza resistance. The total resistance for the interface is therefore the sum of all concentric contributions, yielding the finite width approximation for the interfacial thermal conductance in Eq. \eqref{eq:nonperiodicg}.

\section{Results and Discussion}

\subsection{Thermal Transport Properties}
The solvent thermal conductivity and interfacial thermal conductance values for the planar interfaces are shown in Table \ref{tab:planar_thermal_values}. As expected, solvent thermal conductivity values are similar to those calculated in  pristine  Au(111). However, the values of $G$ are higher here compared to the  pristine  interface.\cite{ss22} In this work, there is no significant difference in $G$ across the planar facets in contact with the ligand, meaning that the surface ordering of gold atoms does not dominate thermal transport in these systems.

\begin{table}[h]
    \centering
    \begin{tabular}{l | l c c c}
    Facet & Metal &  $\lambda~\mathrm{(W m^{-1}~K^{-1})}$ & $G~\mathrm{(MW~m^{-2}~K^{-1})}$ \\ \hline
    \multirow{2}{1.3in}{(111)} & nonpolarizable & 1.02 $\pm$ 0.02 & 211 $\pm$ 15 \\
     & polarizable & 1.01 $\pm$  0.03 &  208 $\pm$ 4 \\ \hline
    \multirow{2}{1.3in}{(110)}& nonpolarizable & 1.01 $\pm$ 0.03 & 215 $\pm$ 7 \\
     & polarizable & 1.01 $\pm$ 0.03 & 217 $\pm$ 8 \\ \hline
     \multirow{2}{1.3in}{(100)} & nonpolarizable & 0.98 $\pm$ 0.04 & 207 $\pm$ 9 \\
    & polarizable &  1.00 $\pm$ 0.03 & 209 $\pm$ 5 \\ 
    \end{tabular}
    \caption{Solvent Thermal Conductivity ($\lambda$) and Interfacial Thermal Conductance ($G$) Values for Systems Containing Planar Gold Interfaces Capped with Thiolated PEG}
    \label{tab:planar_thermal_values}
\end{table}

The interfacial thermal conductance values for the spherical systems are shown in Table \ref{tab:spherical_thermal_values}. When compared with the planar interfaces, the reported mean values of $G$ are larger for systems with higher curvature, but we note that only a few of these values fall outside of the 95\% confidence intervals of comparable systems, i.e. the same treatment of metal polarizability. The differences in the value of $G$ for 10 \AA{} and 20 \AA{} particles fall within confidence intervals. 
%$G$ also appears to be insensitive to metal polarizability, a significant difference from a previous study of citrate-capped spherical systems, where water and ionic species were in direct contact with the metal.\cite{ss22}  Gold capped with a weakly bound ligand such as citrate utilizes direct contact with solvent molecules near the surface to transfer heat across the interface, but a strongly bound ligand such as thiolated PEG will remain in direct contact with the gold, yielding a uniformly dense network of ligands for heat to travel across regardless of metallic morphology. Given the values in Tables \ref{tab:planar_thermal_values} and \ref{tab:spherical_thermal_values}, we can conclude that $G$ in gold capped with thiolated PEG is not significantly affected by metallic facet, metal polarizability, or surface curvature, suggesting that the presence of the strongly bound ligand obscures other variables which have been shown to effect $G$ in systems containing gold that is either bare or capped with a more weakly binding ligand.\cite{baw22, lep23, og22, ss22} 
%The simulations have an average temperature of 300 $\pm$ 30~K.

\begin{table}[h]
    \centering
    \begin{tabular}{l | l c c }
    Radius & Metal & $G~\mathrm{(MW~m^{-2}~K^{-1})}$ \\ \hline
    \multirow{2}{1.3in}{10 \AA{}} & nonpolarizable & 250 $\pm$ 25\\
     & polarizable & 217 $\pm$ 77\\ \hline
    \multirow{2}{1.3in}{20 \AA{}} & nonpolarizable & 307 $\pm$ 112\\
     & polarizable & 244 $\pm$ 45 \\ 
    \end{tabular}
    \caption{Interfacial Thermal Conductance ($G$) Values for Systems Containing Spherical Gold Interfaces Capped with Thiolated PEG}
    \label{tab:spherical_thermal_values}
\end{table}

\begin{figure}[h]
    \centering
    \includegraphics[width=\linewidth]{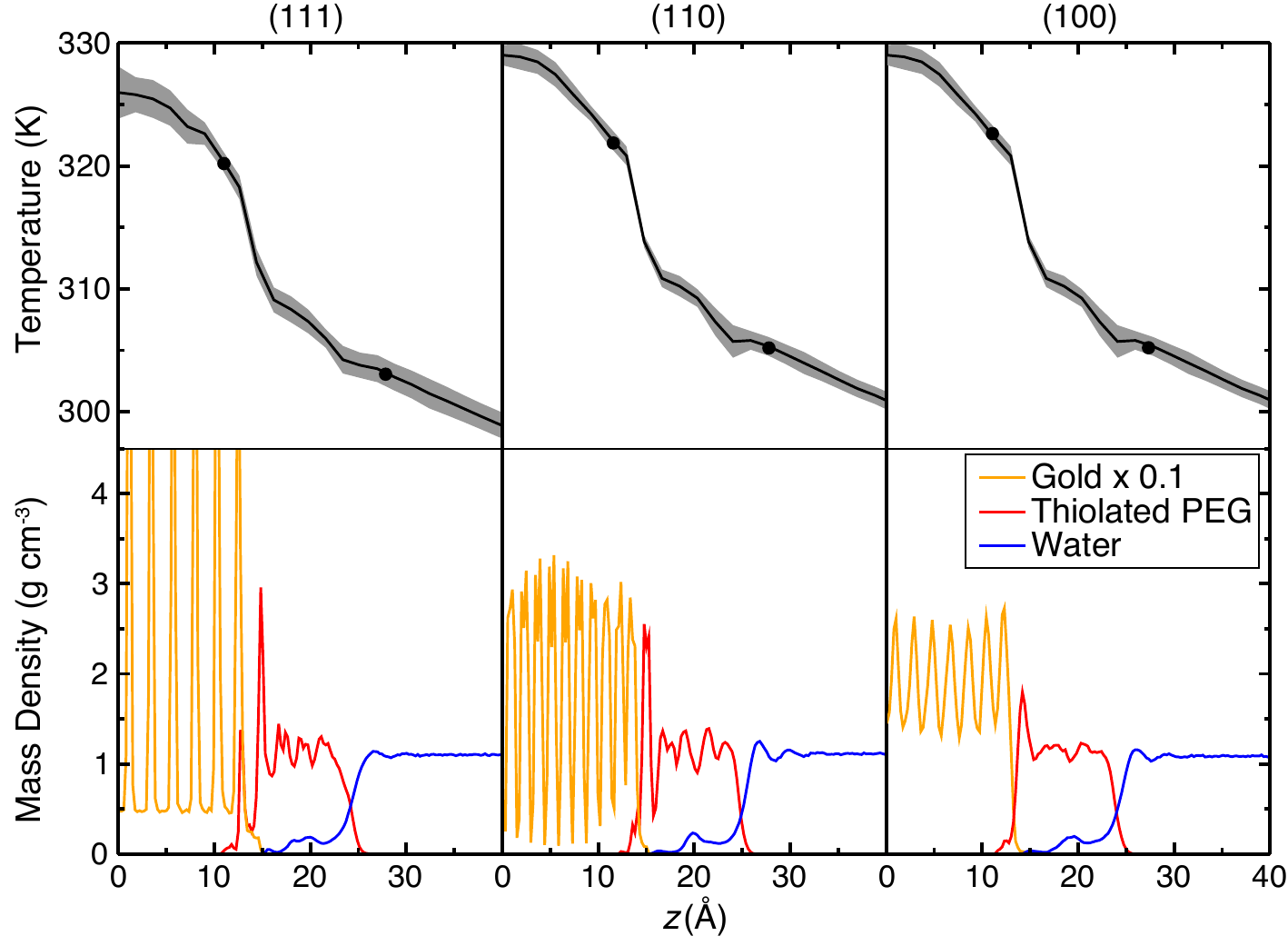}
    \caption{ Under an applied thermal flux, temperature drops are largest at the location of the gold-sulfur bond. (Top) Thermal profiles of planar systems simulated using the nonpolarizable EAM potential. Shading indicates the 95\% confidence interval for computed temperatures, while the black dots are the bounds of the interfacial region used for computing conductance.  (Bottom) Mass densities of each species. $z = 0$ represents the center of the gold slabs.\label{fig:massdensityz}} 
\end{figure}

 To examine the organization of the ligand and solvent layers, we have computed the planar thermal profiles and local mass densities of each species in the systems of interest. These are shown in Figure \ref{fig:massdensityz} (and for the polarizable surface in the Supporting Information). We note that the temperature drops most sharply between the gold surface and the sulfur atoms on the ligand, indicating the location of the highest thermal resistance $\left(R_K = 1 / G \right)$. While ballistic transport can occur across molecular junctions,\cite{hdp16} there is a significant amount of disorder present at this interface.  As can be seen in the mass density profiles, the thiolated PEG molecules partially penetrate the first surface gold layer, embedding sulfur atoms in the top gold plane, and displacing a few surface gold atoms. This penetration occurs to the greatest extent at the Au(111) interfaces. The water molecules also display less ordered layering directly adjacent to the bound ligands in these systems, suggesting that the embedding behavior of the thiolated PEG directly affects the structuring of the solvent. We also note an appreciable density of water that congregates in the spaces between the thiolated PEG molecules. 
% The disorder caused by this interpenetration of sulfur into gold and water into thiolated PEG could explain the lack of ballistic transport across the ligand.  
Other than slight differences in thiolated PEG penetration at Au(111) interfaces, metal polarizability appears to make no significant difference in the ordering and structure of any species.  Thermal profiles and mass densities of each species in planar systems using the polarizable DR-EAM metallic potential are available in the Supporting Information. 

\clearpage

\begin{figure}[h]
    \centering
    \includegraphics[width=\linewidth]{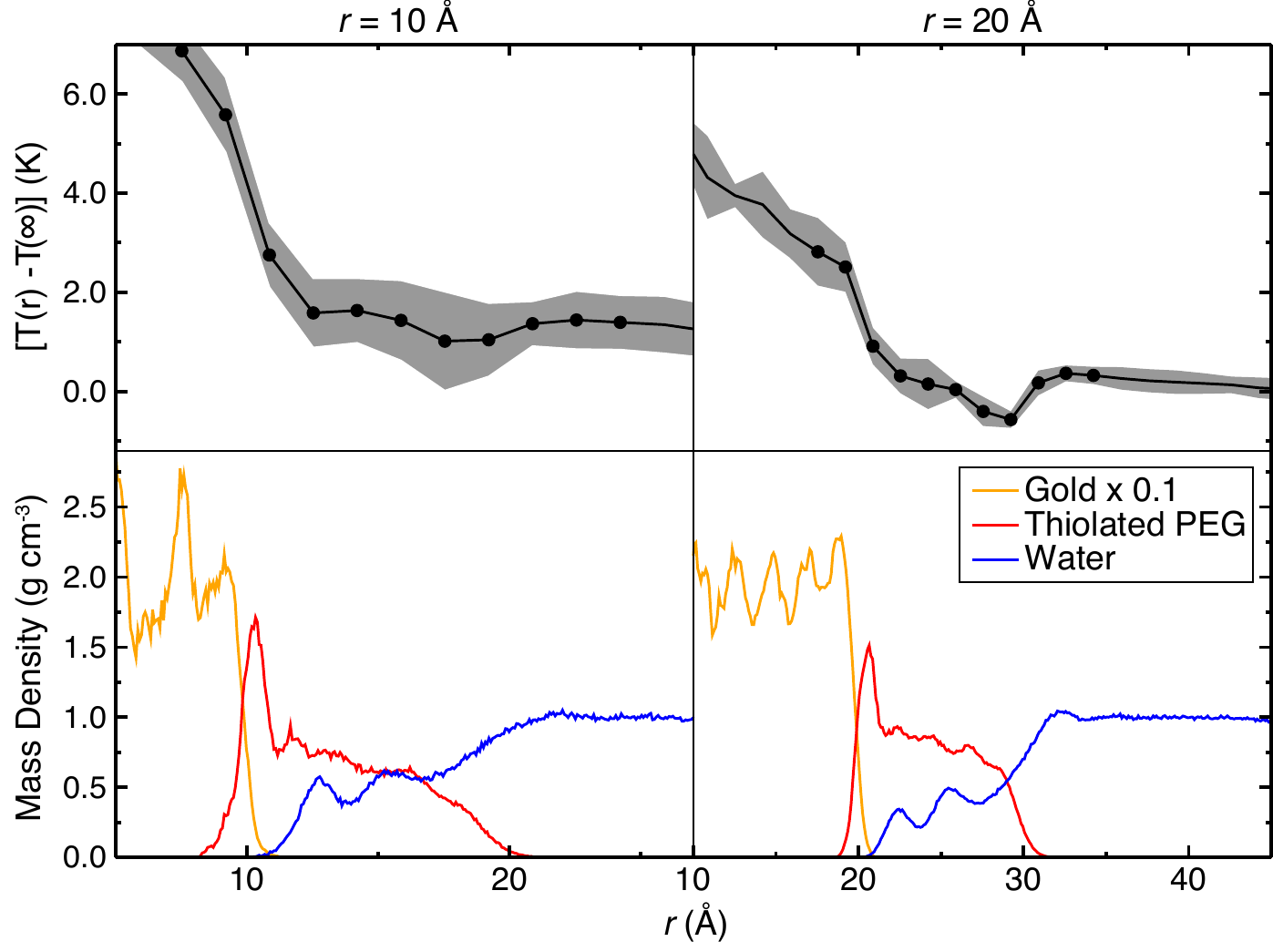}
    \caption{ In the nanospheres, temperature drops under an applied thermal flux are also largest at the Gold-Sulfur bond. (Top) Thermal profiles of nanoparticle systems, relative to the temperature of the solvent far from the particle. The shaded region represents the 95\% confidence intervals. The black data points are the temperature differences in the Kapitza region which were used for calculating conductance.
    (Bottom) Local mass densities of each species for the nanoparticles using the nonpolarizable EAM model. $r = 0$ is at the center of the gold nanosphere. 
    \label{fig:massdensityr}} 
\end{figure}

The  thermal profiles and  local mass densities of all species in the nonpolarizable nanosphere systems are shown in Figure \ref{fig:massdensityr}.  Note that in the nanoparticle systems, average solvent temperatures at the Langevin Hull (far from the particle) spanned a range from 288 to 301 K, so we show here the temperature difference from the solvent near the hull, \textit{i.e.}, as $r \rightarrow \infty$.  When plotted this way, the temperature gaps at the interface are relatively uniform across different samples. The water shows more embedding in the thiolated PEG layer than in planar systems, and more embedding of the ligand layer is also apparent. As in the planar systems, thiolated PEG partially penetrates into the surface of the gold nanospheres. One notable feature is that the water appears to be clustered near the oxygen atoms in the thiolated PEG layer, suggesting that local hydrogen bonding may be important.  The thermal profiles and local mass densities for the polarizable nanosphere systems are available in the Supporting Information. Metal polarizability appears to make no difference in the mass density picture and there are no obvious particle radius effects that would influence the amount of contact between any species.
\clearpage

The large observed interfacial thermal conductance in these systems would require coupling between both the metal surface and the ligand, and between the ligand and solvent. The strong Au-S bonding interaction provides the metal-to-ligand coupling, but what provides the ligand-to-solvent coupling? 
Other groups have commented on the potential link between increased solvent-ligand interactions and an increase in $G$.\cite{hh13, sh16, ks13, ks16} 
To quantify the physical overlap between species, we explored whether molecules in the same physical region (i.e. the ligand layer) had probability of being at the same physical distance from the interface. We calculated the Bhattacharyya coefficient (BC) for each pair of species using normalized mass densities to determine probability densities. In planar systems containing molecules $A$ and $B$, 
\begin{equation}
    BC = \int_{z_L}^{z_R} \sqrt{p_A(z)p_B(z)} \,dz
    \label{eq:bc_planar}
\end{equation}
where $p_A(z)$ and $p_B(z)$ are the probability densities of molecules $A$ and $B$, respectively, calculated using 
\begin{equation}
    p_A(z) = \frac{\rho_A(z)}{\int_{z_L}^{z_R}\rho_A(z) \, dz}
\end{equation}
for species $A$, where $\rho_A(z)$ is the mass density of species A and the bounds of the integrals are the outermost locations of the thiolated PEG in the $z$ direction. Similarly, for spherical systems, 
\begin{equation}
    BC = \int_{r_\mathrm{inner}}^{r_\mathrm{outer}} 4\pi r^{2}  \sqrt{p_A(r)p_B(r)} dr
    \label{eq:bc_spherical}
\end{equation}
where $p_A(r)$ and $p_B(r)$ are the probability densities of molecules $A$ and $B$, respectively. Likewise, in spherical geometries, the probability densities,
\begin{equation}
    p_A(r) = \frac{\rho_A(r)}{\int_{r_\text{inner}}^{r_\text{outer}} 4\pi r^{2} \rho_A(r) dr}
\end{equation}
for species $A$, where $\rho_A(r)$ is the mass density for molecule $A$ and the bounds of the integrals span all values $r$ containing thiolated PEG. The resulting BC values are unitless numbers between 0 and 1, where 0 signifies no overlap between the two species and 1 signifies complete overlap in the ligand region. A higher BC would indicate increased contact between two species. We note that in other investigations of thermal transport at metal / nonmetal interfaces, BCs have been calculated using the vibrational densities of states (VDOS) of two species at an interface,\cite{baw22} while the treatment here provides a measure of physical interpenetration rather than vibrational overlap. However, we have also investigated the vibrational overlap, which is shown in Table \ref{tab:bcp} in the Supporting Information. 

To investigate the extent of contact between the two pairs of molecules, we calculated the corresponding BC values which are shown in Table \ref{tab:bcm}. There is a large amount of overlap between water and thiolated PEG, with little difference across the three planar geometries. Additionally, gold polarizability does not significantly alter the BC in either the planar or spherical cases. Interestingly, the BC values in the spherical systems are higher than those for the planar systems, indicating a significant increase in physical contact between the ligand and solvent. This could explain a mechanism for enhanced thermal transport at curved nanoparticle interfaces (relative to their planar counterparts) with porous ligand layers.

In the planar systems, the density-based BC values confirm that the (111) facet has more embedding of the thiolated PEG into the top gold layer (see Figure \ref{fig:massdensityz}). However, the difference in BC values here is small. Water / thiolated PEG BC values indicate nearly identical overlap across all facets, and this is reflected in the resulting values of $G$. 

\begin{table}[h]
    \centering
    \begin{tabular}{l | l | c  c  }
    \multirow{2}{*}{Facet} & \multirow{2}{*}{Metal} &  \multicolumn{2}{c}{BC with thiolated PEG (unitless)} \\ \cline{3-4}
    & & Water & Gold \\ \toprule
    \multirow{2}{1.3in}{(111)} & nonpolarizable & 0.552 $\pm$ 0.028 & 0.107 $\pm$ 0.032\\
     & polarizable & 0.526 $\pm$ 0.034 & 0.126 $\pm$ 0.026\\ \midrule
    \multirow{2}{1.3in}{(110)}& nonpolarizable & 0.564 $\pm$ 0.025 & 0.048 $\pm$ 0.011\\
     & polarizable & 0.557 $\pm$ 0.038 & 0.050 $\pm$ 0.005\\ \midrule
     \multirow{2}{1.3in}{(100)} & nonpolarizable & 0.559 $\pm$ 0.059 & 0.058 $\pm$ 0.012\\
    & polarizable & 0.554 $\pm$ 0.043 & 0.043 $\pm$ 0.008\\ \midrule
    \multirow{2}{1.3in}{NP ($r = 10$~\AA{})} & nonpolarizable & 0.671 $\pm$ 0.038 & 0.258 $\pm$ 0.037\\
    & polarizable & 0.661 $\pm$ 0.023 & 0.302 $\pm$ 0.031 \\ \midrule
    \multirow{2}{1.3in}{NP ($r = 20$~\AA{})} & nonpolarizable & 0.664 $\pm$ 0.029 & 0.201 $\pm$ 0.052\\
    & polarizable & 0.684 $\pm$ 0.026 & 0.175 $\pm$ 0.095\\ \bottomrule
    \end{tabular}
    \caption{Bhattacharyya Coefficients showing degree of density overlap between the thiolated PEG and the other two components of the metal / ligand / water systems.}
    \label{tab:bcm}
\end{table}

\subsection{Factors Governing Ligand-to-Solvent Coupling}
The strong metal-to-ligand coupling is one reason that $G$ is relatively high in these systems, but the orientation, interpenetration, and hydrogen bonding preferences of the thiolated PEG molecules are all worth investigating, as a previous study has correlated ligand-to-solvent orientational alignment with increased thermal transport.\cite{ks13} We determined the average alignment of ligand molecules relative to the gold surface by calculating the second order Legendre parameter,
\begin{equation}
    \left<P_2\right> = \frac{1}{2} \left<3 \cos^2\theta - 1\right>
    \label{eq:p2}
\end{equation}
where $\theta$ is the angle between the vector between the sulfur and terminal hydroxyl hydrogen atoms in each thiolated PEG molecule and the vector normal to the metal surface at the attachment point. If all thiolated PEG molecules are oriented perpendicular to the metal surface, then $\left<P_2\right> = 1$. Likewise, completely disordered ligands will produce $\left<P_2\right> = 0$. 

\begin{table}[h]
    \centering
    \begin{tabular}{l | l | c | c }
    \multirow{2}{*}{Gold Structure} & \multirow{2}{*}{Metal} &  $\left<P_2\right>$ & H-bonding density \\ 
                                    &                        &  (unitless) &  ($\text{\AA{}}^{-2}$) \\ \hline
\hline
    \multirow{2}{*}{(111)} & nonpolarizable & 0.819 $\pm$ 0.013 & 0.147 $\pm$  0.002\\
                           & polarizable & 0.819 $\pm$ 0.007 & 0.147 $\pm$ 0.007\\ \hline
    \multirow{2}{*}{(110)} & nonpolarizable & 0.836 $\pm$ 0.013 & 0.142 $\pm$ 0.003 \\
                           & polarizable & 0.835 $\pm$ 0.014 & 0.145 $\pm$ 0.005 \\ \hline
    \multirow{2}{*}{(100)} & nonpolarizable & 0.805 $\pm$ 0.009 & 0.144 $\pm$ 0.002 \\
                           & polarizable & 0.812 $\pm$ 0.014 & 0.143 $\pm$ 0.005 \\ \hline
     \multirow{2}{*}{NP ($r$ = 10 \AA{})} & nonpolarizable & 0.709 $\pm$ 0.010 & 0.132 $\pm$ 0.002\\
                                          & polarizable & 0.715 $\pm$ 0.006 & 0.131 $\pm$ 0.001 \\ \hline
    \multirow{2}{*}{NP ($r$ = 20 \AA{})}  & nonpolarizable & 0.743 $\pm$ 0.006 & 0.116 $\pm$ 0.002\\
                                          & polarizable & 0.759 $\pm$ 0.006 & 0.118 $\pm$ 0.002\\ 
    \end{tabular}
    \caption{Structural properties of the planar and nanosphere interfaces. This table provides second order Legendre parameters and Hydrogen bond surface densities.}
    \label{tab:structural}
\end{table}

The orientational order parameters for all systems are shown in Table \ref{tab:structural}.  Within confidence intervals, the planar interfaces exhibit  ligand chain ordering that are independent of facet, while high curvature nanospheres ($r = 10$~\AA{}) show the lowest orientational ordering.  The lower curvature nanospheres ($r = 20$~\AA{}) are intermediate.  We note that there is a slight preference for ligand ordering on the (110) facet relative to the other planar facets. Most of the orientational ordering parameters for nonpolarizable planar systems are within error of their polarizable counterparts. There is a significant increase in ligand ordering in the larger nanospheres when the gold is polarizable, although this is the only system that exhibits this behavior. The lower orientational ordering in the nanosphere systems is to be expected, as the configurational volume available to the terminal ends of the chains is significantly higher than in the planar systems (with the same surface packing density).

\begin{figure}[h!]
    \centering
    \includegraphics[width=\linewidth]{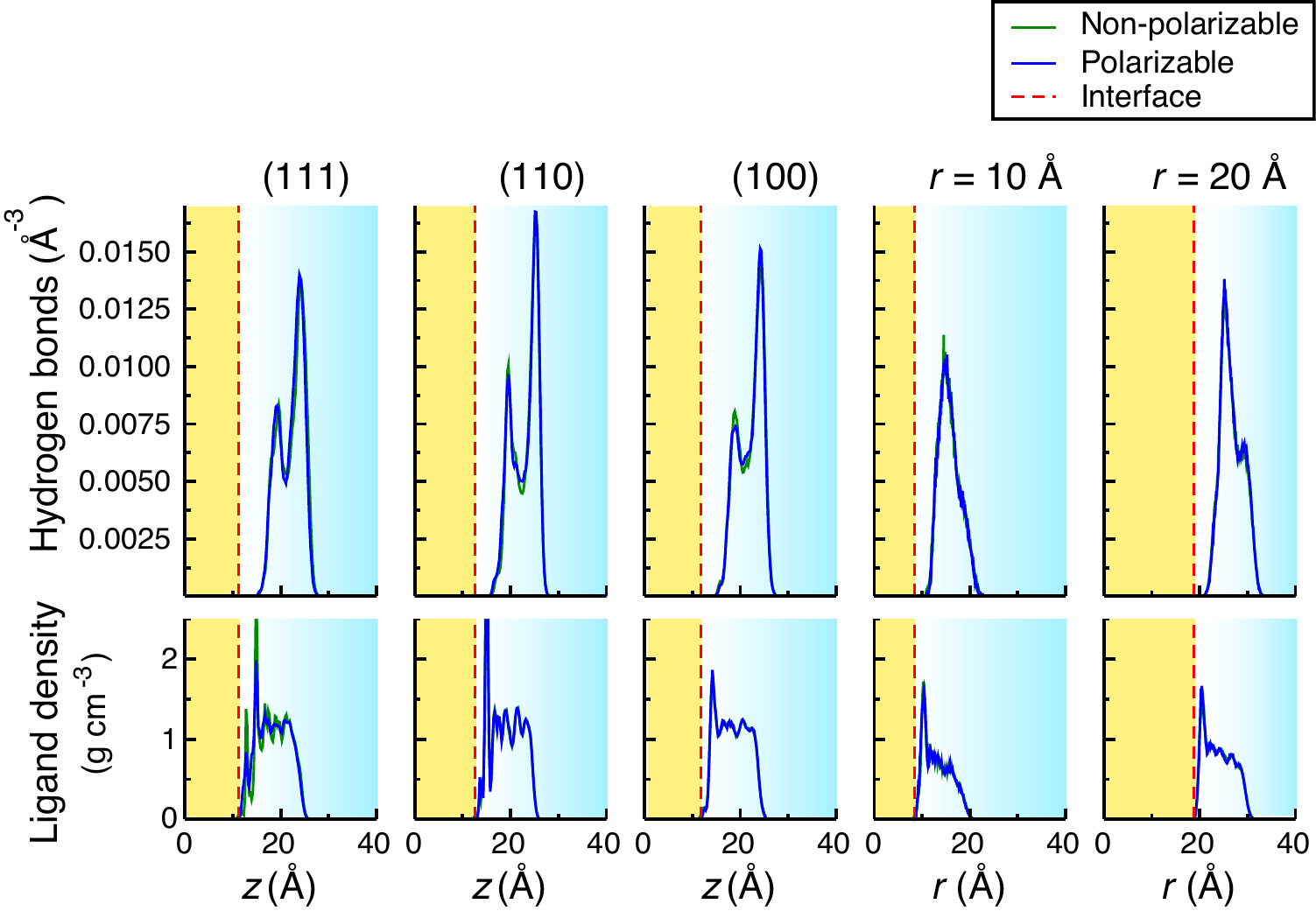}
    \caption{ Hydrogen bonding densities (Top) and ligand mass densities (Bottom) for all interfaces,  illustrating the changes in the network of hydrogen bonding between planar and nanoparticle systems. Metal polarizability does not significantly change the hydrogen bonding densities, while the shape and size of the gold particles appear to have an impact. The planar systems exhibit the most hydrogen bonding involving the terminal oxygen, while hydrogen bonding with the ether oxygen is most prevalent in the nanospheres. Gold surface curvature affects the configurational freedom in thiolated PEG chains, and determines how many water molecules can come within the hydrogen bonding distance of the ligand. \label{fig:hbondzrvol}} 
\end{figure}

To further probe the ligand-solvent interactions, we determined the hydrogen bonding density between water and thiolated PEG. We examined all potential donor and acceptor atoms (including the hydroxyl hydrogen) inside the hydrogen bonding distance (3.5 \AA{}) from water, with an HOH angle cutoff of 30\degree. Local hydrogen bond densities were computed as a function of surface normal coordinate ($z$ in the planar systems, $r$ in the nanospheres). Hydrogen bond densities  and ligand mass densities  are shown in Figure \ref{fig:hbondzrvol},  and the integrated hydrogen bond densities are shown in Table \ref{tab:structural}. These values provide a density of ligand-to-solvent hydrogen bonds per surface area of the metal.  For planar systems
\begin{equation}
    \text{H-bonding density per \AA{}$^2$} =  \int_{0}^{z} \rho(z) dz
\end{equation}
where $\rho(z)$ is the hydrogen bonding per \AA{}$^3$.
For spherical systems, integration is done in spherical coordinates
\begin{equation}
    \text{H-bonding density per \AA{}$^2$} = \int_{0}^{\infty} \left(\frac{r}{R}\right)^2 \rho(r)dr
\end{equation}
where $\rho(r)$ is the hydrogen bonding per \AA{}$^3$ and $R$ is the radius of the gold nanosphere.

\clearpage

From these data, we can determine which atoms in the thiolated PEG ligand are responsible for hydrogen bonding, and how the magnitude of hydrogen bonding varies as a function of metal polarizability, facet, shape, and size. An extensive hydrogen bonding network between thiolated PEG and water would mean an additional layer of contact between the ligand and solvent, possibly explaining the high $G$ values.

In Figure \ref{fig:hbondzrvol}, the planar systems each display two distinct peaks, where each peak represents hydrogen bonding at one of the oxygen atoms in the ligand. The outermost peak comes from the terminal hydroxyl group. The other, largely based on spacing between the peaks, is due to the ether oxygen that is three atoms closer to the surface, while the oxygen closest to sulfur does not participate in hydrogen bonding. In the planar systems, the majority of hydrogen bonding occurs at the terminal oxygen. Due to the densely packed network of thiolated PEG molecules, fewer water molecules are able to congregate between them to facilitate hydrogen bonding. The magnitude of the hydrogen bonding densities are similar across all planar geometries.

For spherical systems, hydrogen bonding with water favors one of the ether oxygen atoms, although the terminal hydroxyl group also participates. As in the planar systems, the ether oxygen most likely to form hydrogen bonds is the outermost one. It is also useful to note that the total hydrogen bonding per unit area of the gold surface is reduced in the nanospheres (see Table \ref{tab:structural}).  We observe a higher degree of ligand-to-water hydrogen bonds in the planar systems, but the hydrogen bonding is deeper in the ligand layer for the nanospheres. Polarizability does not appear to significantly affect the magnitude or distribution of hydrogen bonding for any system.

Additionally, we computed charge densities as a function of the distance from the interfaces. Although there are some observable differences between facets and between the planar systems and nanospheres, particularly on the thiolated PEG side of the interface, the polarizability model appears to have little effect on the interfacial thermal conductance in these systems.  Plots of charge densities are provided in the Supporting Information.

\subsection{Analysis of Vibrational Power Spectra}
The models most widely used to understand interfacial thermal transport have their origins in the diffuse mismatch model (DMM).\cite{ets89}  Although it is rarely quantitative  and does not consider interactions between adjoining materials,  the DMM can point to factors responsible for enhanced (or diminished) heat transfer. It assumes that phonons that strike the interface are elastically scattered and transferred into the adjoining material with a transmission probability that depends on
the phonon densities of states (DOS) in the two materials. At the interface between materials $a$ and $b$, the interfacial thermal conductance
\begin{equation}
    G_{ab} = \frac{1}{4\pi}\sum_{p} \int_{0}^{\pi/2} \sin{\theta} d\theta \int_{0}^{2 \pi} d\phi \int_0^\infty d \omega ~ \left( \hbar \omega \frac{\partial f}{\partial T} \nu_a \rho_a(\omega) \tau_{ab} \cos{\theta} \right)
    \label{dmm}
\end{equation}
where $f$ is the Bose-Einstein function 
\begin{equation}
f(\omega, T) = \frac{1}{(e^{\hbar \omega / k_B T} - 1)},
\end{equation}
and phonons with a polarization $p$, group velocity $\nu_a(\omega, p)$, and incident angles $\theta$ and $\phi$ are transmitted between $a$ and $b$ with a transmission probability $\tau_{ab}(\omega)$.\cite{cm16} 

If we assume that the transmission probability and group velocity do not depend on the incident scattering angle, the angular integrals are easily solved.  Additionally, the derivative of the Bose-Einstein function has the effect of weighting the lowest frequency portion of the VDOS. One of the simplest models for the transmission probability involves using detailed balance to connect forward and reverse scattering at a frequency $\omega$, that is,  $\tau_{ab}(\omega) = 1 - \tau_{ba}(\omega)$.  The connection between $G_{ab}$ and $G_{ba}$ is made clear by treating the remaining contribution to interfacial thermal conductance in a symmetric form,
\begin{equation}
    \rho_a(\omega) \tau_{ab}(\omega) \approx \frac{\rho_a(\omega) \rho_b(\omega)}{\rho_a(\omega) + \rho_b(\omega)}.
\end{equation}
This shows the importance of having the same phonon frequencies present on both sides of the interface  to carry heat between $a$ and $b$ (and between $b$ and $a$). This portion of the integrand survives only if both materials have contributions at the same frequencies.
We note that while the DMM has limited success as a predictive theory, we can still use it to interpret features in our vibrational power spectra that may explain the higher mean $G$ values in the nanospheres.

To explore the roles of particle curvature, internal hydrogen bonding and metal polarizability, we have computed the power spectra for vibrational motion,
\begin{equation}
    \rho(\omega) = \frac{1}{\sqrt{2\pi}} \int_{-\infty}^{\infty} \langle \mathbf{v}(t) \cdot \mathbf{v}(0) \rangle e^{-i\omega t} \, dt
\end{equation} 
where the angle bracket averages over all particles of a given type. We divided the systems into four regions: interfacial gold (within 5 \AA{} of the interface), the ligand itself, interfacial solvent that is embedded in the thiolated PEG layer, and bulk solvent that lies outside of the ligand layer, for both the planar and nanoparticle systems. Each system was simulated for an additional 50 ps in the NVE ensemble for the planar systems and in a Langevin Hull for the nanoparticle systems. Velocity autocorrelation functions with a time granularity of 3 fs were calculated from these trajectories. The power spectra are discrete Fourier transforms of these autocorrelation functions. Power spectra were normalized so that they integrate to unity over the entire frequency range. 

We find that the power spectra do not depend on the polarizability model used to represent the metal, and we also observe very similar spectra for all three planar metal facets. Similarly, the power spectra for the two nanosphere systems are nearly identical. To aid in comparison, we have computed the spectra, averaged across metal facets (for the planar systems) and polarizability. For spherical systems, the spectra were averaged for particles of different radii and polarizability. For completeness, the spectra for all components of the individual systems can be found in the Supporting Information (Figures \ref{fig:planarpspect} and \ref{fig:nppspect}).

\begin{figure}[h]
    \centering
    \includegraphics[width=\linewidth]{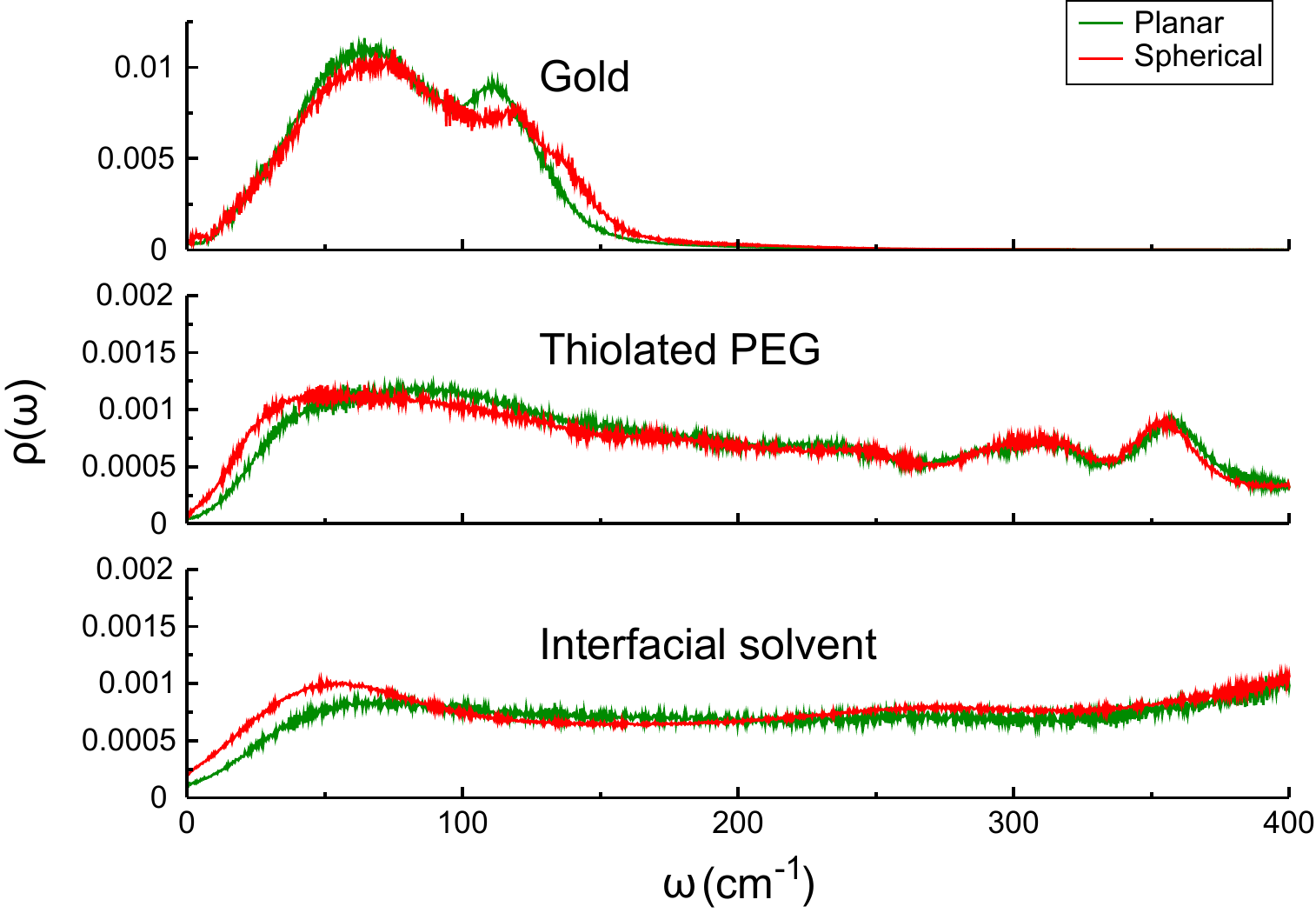}
    \caption{Averaged power spectra for gold, thiolated PEG and the interfacial solvent in planar and nanosphere systems. We show here only the moderate frequency portion ($0 - 400 \mathrm{~cm}^{-1}$). Lower-frequency modes attributed to collective motion are most important for heat transfer, and we observe a slightly higher population from $0 \text{ to } 80 \mathrm{~cm}^{-1}$ in the interfacial layers attached to the nanospheres. This difference persists in the interfacial solvent that is trapped in the thiolated PEG layer. Note that the planar curve is an average of all polarizable and nonpolarizable simulations of (111), (110), and (100) facets, and the spherical curve is an average of all polarizable and nonpolarizable simulations of nanospheres with radii of 10 and 20 \AA{}.
    \label{fig:thiolpegwaterpspect}} 
\end{figure}

The averaged power spectra for gold, thiolated PEG, and the interfacial solvent are shown in Figure \ref{fig:thiolpegwaterpspect}. In the gold nanospheres, we observe a broadening and shift in the higher frequency peak that has been previously attributed to undercoordinated gold atoms at the surfaces of the spheres.\cite{ast17, mj22}  We also observe an increase in the low frequency heat carrying modes (0 - 70 $\text{cm}^{-1}$) for the thiolated PEG. Because of the Bose-Einstein weighting of lower frequency modes, the enhanced population at these frequencies may explain the higher mean $G$ values found in the nanospheres. In the interfacial solvent, there is again an increased population at low frequencies for the spherical systems, here between 0 to 80 $\text{cm}^{-1}$.  We also note the lack of overlap between the thiolated PEG and gold spectra at low frequencies which potentially explains the larger temperature drop between the two species. 

We have also tabulated the Bhattacharyya coefficients (BC) from vibrational power spectra for the same pairs of species as in Table \ref{tab:bcm}, which are available in the Supporting Information. Here, the integral range spans all computed frequencies (0 - 6000 $\text{cm}^{-1}$). While the mean BC value for the gold / thiolated PEG pair is higher in the 10 \AA{} nanospheres, the overlap between all other pairs is nearly the same for all systems, likely due to the unbiased frequency weighting in the calculation of these coefficients.  

\clearpage

\section{Conclusions}
We have studied the effects of metal polarizability, shape, facet, and size on heat transfer processes in gold interfaces functionalized with a low molecular weight thiolated PEG. We found that the mean values of $G$ in spherical systems are generally higher than those in planar systems. We have also determined that metal polarizability and the choice of planar facet make little difference in heat transfer processes in these systems. Additionally, the solvent thermal conductivity remains nearly constant for all planar systems with the same ligand grafting density.  From thermal profiles, we determined that the largest temperature drop occurs between the gold surface and sulfur atoms in thiolated PEG, meaning that thermal resistivity is highest at this portion of the interface. 

We observed an overlap in the mass densities of thiolated PEG and gold, meaning that the ligands are capable of penetrating and disrupting the gold surface. From the corresponding BCs, we determined that there is an increased level of physical contact between all adjacent pairs of species in the nanospheres. We also found that there is an enhanced population of low frequency heat-carrying modes in the vibrational power spectra of the thiolated PEG and interfacial solvent in the nanospheres, and this may contribute to the higher mean $G$ values in those systems. The thiolated PEG molecules are less ordered in the spherical systems, with a larger configurational volume available for the terminal ends of the ligands. The ether oxygen in thiolated PEG participates in hydrogen bonding with the solvent to a greater degree than the terminal hydroxyl group here, a trend that is reversed in planar systems. Interestingly, there is an overall lower hydrogen bonding density in the nanospheres than in planar systems. Our results suggest that the strong metal-to-ligand coupling present in thiolated PEG on gold is important for interfacial thermal conductance.  This appears to dominate the effects of the metal polarizability, facet, and surface curvature.  There may be also be a weak dependence on the location and kind of ligand-to-solvent hydrogen bonding.

Thiolated PEG is a biocompatible non-ionic ligand which allows gold nanoparticles to be solvated in water. It may not be a surprise that metal polarizability is relatively unimportant for interfacial thermal conductance through a nonionic protecting group, but there are other bio-compatible, aqueous, \textit{ionic} ligands, such as cetyltrimethylammonium bromide (CTAB), which are candidates for protecting nanospheres in  photothermal therapies. CTAB also has a different binding affinity to gold than thiolated PEG, and it prevents gold aggregation through a different mechanism (i.e., through the formation of micelles). Studies of other biocompatible ligands would help us better understand how heat transfer is affected by the same factors studied here, as well as how the ionic character of such a ligand would affect interfacial thermal conductance.

\begin{acknowledgement}
Support for this project was provided by the National Science Foundation under grant CHE-1954648. Computational time was provided by the Center for Research Computing (CRC) at the University of Notre Dame.
The authors would also like to thank Benjamin M. Harless and C.R. Drisko for helpful discussions regarding data analysis.
\end{acknowledgement}

\begin{suppinfo}
Details regarding system composition, force field parameters, equilibration,  method validation,  data collection, additional vibrational power spectra,  thermal profiles, mass densities,  and charge densities described in this work.
\end{suppinfo}

\newpage
\bibliography{ThiolPEG}

\providecommand{\latin}[1]{#1}
\makeatletter
\providecommand{\doi}
  {\begingroup\let\do\@makeother\dospecials
  \catcode`\{=1 \catcode`\}=2 \doi@aux}
\providecommand{\doi@aux}[1]{\endgroup\texttt{#1}}
\makeatother
\providecommand*\mcitethebibliography{\thebibliography}
\csname @ifundefined\endcsname{endmcitethebibliography}
  {\let\endmcitethebibliography\endthebibliography}{}
\begin{mcitethebibliography}{58}
\providecommand*\natexlab[1]{#1}
\providecommand*\mciteSetBstSublistMode[1]{}
\providecommand*\mciteSetBstMaxWidthForm[2]{}
\providecommand*\mciteBstWouldAddEndPuncttrue
  {\def\EndOfBibitem{\unskip.}}
\providecommand*\mciteBstWouldAddEndPunctfalse
  {\let\EndOfBibitem\relax}
\providecommand*\mciteSetBstMidEndSepPunct[3]{}
\providecommand*\mciteSetBstSublistLabelBeginEnd[3]{}
\providecommand*\EndOfBibitem{}
\mciteSetBstSublistMode{f}
\mciteSetBstMaxWidthForm{subitem}{(\alph{mcitesubitemcount})}
\mciteSetBstSublistLabelBeginEnd
  {\mcitemaxwidthsubitemform\space}
  {\relax}
  {\relax}

\bibitem[Vines \latin{et~al.}(2019)Vines, Yoon, Ryu, Lim, and Park]{jbv19}
Vines,~J.~B.; Yoon,~J.-H.; Ryu,~N.-E.; Lim,~D.-J.; Park,~H. Gold Nanoparticles
  for Photothermal Cancer Therapy. \emph{Front. Chem.} \textbf{2019}, \emph{7},
  1--16\relax
\mciteBstWouldAddEndPuncttrue
\mciteSetBstMidEndSepPunct{\mcitedefaultmidpunct}
{\mcitedefaultendpunct}{\mcitedefaultseppunct}\relax
\EndOfBibitem
\bibitem[Huang and El-Sayed(2010)Huang, and El-Sayed]{xh10}
Huang,~X.; El-Sayed,~M.~A. Gold nanoparticles: Optical properties and
  implementations in cancer diagnosis and photothermal therapy. \emph{J. Adv.
  Res.} \textbf{2010}, \emph{1}, 13--28\relax
\mciteBstWouldAddEndPuncttrue
\mciteSetBstMidEndSepPunct{\mcitedefaultmidpunct}
{\mcitedefaultendpunct}{\mcitedefaultseppunct}\relax
\EndOfBibitem
\bibitem[Bai \latin{et~al.}(2020)Bai, Wang, Song, Feng, Chen, Zhang, and
  Feng]{xb20}
Bai,~X.; Wang,~Y.; Song,~Z.; Feng,~Y.; Chen,~Y.; Zhang,~D.; Feng,~L. The Basic
  Properties of Gold Nanoparticles and their Applications in Tumor Diagnosis
  and Treatment. \emph{Int. J. Mol. Sci.} \textbf{2020}, \emph{21}, 2480\relax
\mciteBstWouldAddEndPuncttrue
\mciteSetBstMidEndSepPunct{\mcitedefaultmidpunct}
{\mcitedefaultendpunct}{\mcitedefaultseppunct}\relax
\EndOfBibitem
\bibitem[Yang \latin{et~al.}(2019)Yang, Liang, Ma, Wang, and Huang]{wy19}
Yang,~W.; Liang,~H.; Ma,~S.; Wang,~D.; Huang,~J. Gold nanoparticle based
  photothermal therapy: Development and application for effective cancer
  treatment. \emph{Sustain. Mater. Technol.} \textbf{2019}, \emph{22},
  1--29\relax
\mciteBstWouldAddEndPuncttrue
\mciteSetBstMidEndSepPunct{\mcitedefaultmidpunct}
{\mcitedefaultendpunct}{\mcitedefaultseppunct}\relax
\EndOfBibitem
\bibitem[Sztandera \latin{et~al.}(2019)Sztandera, Gorzkiewicz, and
  Klajnert-Maculewicz]{ks19}
Sztandera,~K.; Gorzkiewicz,~M.; Klajnert-Maculewicz,~B. Gold nanoparticles in
  cancer treatment. \emph{Mol. Pharmaceutics} \textbf{2019}, \emph{16},
  1--23\relax
\mciteBstWouldAddEndPuncttrue
\mciteSetBstMidEndSepPunct{\mcitedefaultmidpunct}
{\mcitedefaultendpunct}{\mcitedefaultseppunct}\relax
\EndOfBibitem
\bibitem[Lim \latin{et~al.}(2011)Lim, Li, Ng, Yung, and Bay]{zjl11}
Lim,~Z.-Z.~J.; Li,~J.-E.~J.; Ng,~C.-T.; Yung,~L.-Y.~L.; Bay,~B.-H. Gold
  nanoparticles in cancer therapy. \emph{Acta Pharmacol. Sin.} \textbf{2011},
  \emph{32}, 983--990\relax
\mciteBstWouldAddEndPuncttrue
\mciteSetBstMidEndSepPunct{\mcitedefaultmidpunct}
{\mcitedefaultendpunct}{\mcitedefaultseppunct}\relax
\EndOfBibitem
\bibitem[Poole(2015)]{lbp15}
Poole,~L.~B. The basics of thiols and cysteines in redox biology and chemistry.
  \emph{Free Radic. Biol. Med.} \textbf{2015}, \emph{80}, 148--157\relax
\mciteBstWouldAddEndPuncttrue
\mciteSetBstMidEndSepPunct{\mcitedefaultmidpunct}
{\mcitedefaultendpunct}{\mcitedefaultseppunct}\relax
\EndOfBibitem
\bibitem[Mirkin \latin{et~al.}(1996)Mirkin, Letsinger, Mucic, and
  Storhoff]{cam96}
Mirkin,~C.~A.; Letsinger,~R.~L.; Mucic,~R.~C.; Storhoff,~J.~J. A DNA-based
  method for rationally assembling nanoparticles into macroscopic materials.
  \emph{Nature} \textbf{1996}, \emph{382}, 1--23\relax
\mciteBstWouldAddEndPuncttrue
\mciteSetBstMidEndSepPunct{\mcitedefaultmidpunct}
{\mcitedefaultendpunct}{\mcitedefaultseppunct}\relax
\EndOfBibitem
\bibitem[Zhang \latin{et~al.}(2012)Zhang, Servos, and Liu]{xz12}
Zhang,~X.; Servos,~M.~R.; Liu,~J. Instantaneous and Quantitative
  Functionalization of Gold Nanoparticles with Thiolated DNA Using a
  pH-Assisted and Surfactant-Free Route. \emph{J. Am. Chem. Soc.}
  \textbf{2012}, \emph{134}, 7266--7269\relax
\mciteBstWouldAddEndPuncttrue
\mciteSetBstMidEndSepPunct{\mcitedefaultmidpunct}
{\mcitedefaultendpunct}{\mcitedefaultseppunct}\relax
\EndOfBibitem
\bibitem[Peng \latin{et~al.}(2020)Peng, Yu, and Zheng]{cp20}
Peng,~C.; Yu,~M.; Zheng,~J. In Situ Ligand-Directed Growth of Gold
  Nanoparticles in Biological Tissues. \emph{Nano Lett.} \textbf{2020},
  \emph{20}, 1378--1382\relax
\mciteBstWouldAddEndPuncttrue
\mciteSetBstMidEndSepPunct{\mcitedefaultmidpunct}
{\mcitedefaultendpunct}{\mcitedefaultseppunct}\relax
\EndOfBibitem
\bibitem[Sibuyi \latin{et~al.}(2021)Sibuyi, Moabelo, Fadaka, Meyer, Onani,
  Madiehe, and Meyer]{nrss21}
Sibuyi,~N. R.~S.; Moabelo,~K.~L.; Fadaka,~A.~O.; Meyer,~S.; Onani,~M.~O.;
  Madiehe,~A.~M.; Meyer,~M. Multifunctional Gold Nanoparticles for Improved
  Diagnostic and Therapeutic Applications: A Review. \emph{Nanoscale Res.
  Lett.} \textbf{2021}, \emph{16}, 1--27\relax
\mciteBstWouldAddEndPuncttrue
\mciteSetBstMidEndSepPunct{\mcitedefaultmidpunct}
{\mcitedefaultendpunct}{\mcitedefaultseppunct}\relax
\EndOfBibitem
\bibitem[Lee \latin{et~al.}(2007)Lee, Soonsanga, and Helmann]{jl07}
Lee,~J.-W.; Soonsanga,~S.; Helmann,~J.~D. A complex thiolate switch regulates
  the Bacillussubtilisorganic peroxide sensor OhrR. \emph{PNAS} \textbf{2007},
  \emph{4}, 8743--8748\relax
\mciteBstWouldAddEndPuncttrue
\mciteSetBstMidEndSepPunct{\mcitedefaultmidpunct}
{\mcitedefaultendpunct}{\mcitedefaultseppunct}\relax
\EndOfBibitem
\bibitem[Felice and Selloni(2004)Felice, and Selloni]{rdf04}
Felice,~R.~D.; Selloni,~A. Adsorption modes of cysteine on Au(111): Thiolate,
  amino-thiolate, disulfide. \emph{J. Chem. Phys.} \textbf{2004}, \emph{120},
  4906--4914\relax
\mciteBstWouldAddEndPuncttrue
\mciteSetBstMidEndSepPunct{\mcitedefaultmidpunct}
{\mcitedefaultendpunct}{\mcitedefaultseppunct}\relax
\EndOfBibitem
\bibitem[Yao and Huang(2018)Yao, and Huang]{gy18}
Yao,~G.; Huang,~Q. DFT and SERS Study of L‑Cysteine Adsorption on the Surface
  of Gold Nanoparticles. \emph{J. Phys. Chem. C} \textbf{2018}, \emph{122},
  15241--15251\relax
\mciteBstWouldAddEndPuncttrue
\mciteSetBstMidEndSepPunct{\mcitedefaultmidpunct}
{\mcitedefaultendpunct}{\mcitedefaultseppunct}\relax
\EndOfBibitem
\bibitem[Monti \latin{et~al.}(2016)Monti, Carravetta, and Ågren]{sm16}
Monti,~S.; Carravetta,~V.; Ågren,~H. Decoration of gold nanoparticles with
  cysteine in solution: reactive molecular dynamics simulations.
  \emph{Nanoscale} \textbf{2016}, \emph{8}, 12929--12938\relax
\mciteBstWouldAddEndPuncttrue
\mciteSetBstMidEndSepPunct{\mcitedefaultmidpunct}
{\mcitedefaultendpunct}{\mcitedefaultseppunct}\relax
\EndOfBibitem
\bibitem[Li \latin{et~al.}(2006)Li, Duan, Liu, and Du]{zpl06}
Li,~Z.~P.; Duan,~X.~R.; Liu,~C.~H.; Du,~B.~A. Selective determination of
  cysteine by resonance light scattering technique based on self-assembly of
  gold nanoparticles. \emph{Anal. Biochem.} \textbf{2006}, \emph{351},
  18--25\relax
\mciteBstWouldAddEndPuncttrue
\mciteSetBstMidEndSepPunct{\mcitedefaultmidpunct}
{\mcitedefaultendpunct}{\mcitedefaultseppunct}\relax
\EndOfBibitem
\bibitem[Matthiesen \latin{et~al.}(2012)Matthiesen, Jose, Sorensen, and
  Klabunde]{jem12}
Matthiesen,~J.~E.; Jose,~D.; Sorensen,~C.~M.; Klabunde,~K.~J. Loss of Hydrogen
  upon Exposure of Thiol to Gold Clusters at Low Temperature. \emph{J. Am.
  Chem. Soc.} \textbf{2012}, \emph{134}, 9376--9379\relax
\mciteBstWouldAddEndPuncttrue
\mciteSetBstMidEndSepPunct{\mcitedefaultmidpunct}
{\mcitedefaultendpunct}{\mcitedefaultseppunct}\relax
\EndOfBibitem
\bibitem[Woehrle \latin{et~al.}(2005)Woehrle, Brown, and Hutchison]{ghw05}
Woehrle,~G.~H.; Brown,~L.~O.; Hutchison,~J.~E. Thiol-Functionalized, 1.5-nm
  Gold Nanoparticles through Ligand Exchange Reactions: Scope and Mechanism of
  Ligand Exchange. \emph{J. Am. Chem. Soc.} \textbf{2005}, \emph{127},
  2172--2183\relax
\mciteBstWouldAddEndPuncttrue
\mciteSetBstMidEndSepPunct{\mcitedefaultmidpunct}
{\mcitedefaultendpunct}{\mcitedefaultseppunct}\relax
\EndOfBibitem
\bibitem[Brust \latin{et~al.}(1994)Brust, Walker, Bethell, Schiffrin, and
  Whyman]{mb94}
Brust,~M.; Walker,~M.; Bethell,~D.; Schiffrin,~D.~J.; Whyman,~R. Synthesis of
  Thiol-derivatised Gold Nanoparticles in a Two-phase Liquid-Liquid System.
  \emph{J. Chem. Soc. Commun.} \textbf{1994}, \emph{7}, 801--802\relax
\mciteBstWouldAddEndPuncttrue
\mciteSetBstMidEndSepPunct{\mcitedefaultmidpunct}
{\mcitedefaultendpunct}{\mcitedefaultseppunct}\relax
\EndOfBibitem
\bibitem[Zalipsky(1995)]{sz95}
Zalipsky,~S. Chemistry of polyethylene glycol conjugates with biologically
  active molecules. \emph{Adv. Drug Deliv. Rev.} \textbf{1995}, \emph{16},
  157--182\relax
\mciteBstWouldAddEndPuncttrue
\mciteSetBstMidEndSepPunct{\mcitedefaultmidpunct}
{\mcitedefaultendpunct}{\mcitedefaultseppunct}\relax
\EndOfBibitem
\bibitem[Kou \latin{et~al.}(2014)Kou, Wang, Yuan, Chen, Zhi, Gao, Wang, Guo,
  Xue, Cao, and Guo]{zk14}
Kou,~Z.; Wang,~X.; Yuan,~R.; Chen,~H.; Zhi,~Q.; Gao,~L.; Wang,~B.; Guo,~Z.;
  Xue,~X.; Cao,~W. \latin{et~al.}  A promising gene delivery system developed
  from PEGylated MoS$_2$ nanosheets for gene therapy. \emph{Nanoscale Res.
  Lett.} \textbf{2014}, \emph{9}, 587\relax
\mciteBstWouldAddEndPuncttrue
\mciteSetBstMidEndSepPunct{\mcitedefaultmidpunct}
{\mcitedefaultendpunct}{\mcitedefaultseppunct}\relax
\EndOfBibitem
\bibitem[Feng \latin{et~al.}(2013)Feng, Yang, Shi, Tan, Peng, Wang, and
  Liu]{lf13}
Feng,~L.; Yang,~X.; Shi,~X.; Tan,~X.; Peng,~R.; Wang,~J.; Liu,~Z. Polyethylene
  Glycol and Polyethylenimine DualFunctionalized Nano-Graphene Oxide for
  Photothermally Enhanced Gene Delivery. \emph{Small} \textbf{2013}, \emph{9},
  1989--1997\relax
\mciteBstWouldAddEndPuncttrue
\mciteSetBstMidEndSepPunct{\mcitedefaultmidpunct}
{\mcitedefaultendpunct}{\mcitedefaultseppunct}\relax
\EndOfBibitem
\bibitem[Chen \latin{et~al.}(2018)Chen, Chen, Chen, Li, Chen, Tang, Xie, Luo,
  Wang, Liang, and Yu]{lc18}
Chen,~L.; Chen,~C.; Chen,~W.; Li,~K.; Chen,~X.; Tang,~X.; Xie,~G.; Luo,~X.;
  Wang,~X.; Liang,~H. \latin{et~al.}  Biodegradable Black Phosphorus Nanosheets
  Mediate Specific Delivery of hTERT siRNA for Synergistic Cancer Therapy.
  \emph{ACS Appl. Mater. Interfaces} \textbf{2018}, \emph{10},
  21137--21148\relax
\mciteBstWouldAddEndPuncttrue
\mciteSetBstMidEndSepPunct{\mcitedefaultmidpunct}
{\mcitedefaultendpunct}{\mcitedefaultseppunct}\relax
\EndOfBibitem
\bibitem[Harikrishna \latin{et~al.}(2013)Harikrishna, Ducker, and
  Huxtable]{hh13}
Harikrishna,~H.; Ducker,~W.~A.; Huxtable,~S.~T. The influence of interface
  bonding on thermal transport through solid-liquid interfaces. \emph{Appl.
  Phys. Lett.} \textbf{2013}, \emph{102}, 251606\relax
\mciteBstWouldAddEndPuncttrue
\mciteSetBstMidEndSepPunct{\mcitedefaultmidpunct}
{\mcitedefaultendpunct}{\mcitedefaultseppunct}\relax
\EndOfBibitem
\bibitem[Stocker and Gezelter(2013)Stocker, and Gezelter]{ks13}
Stocker,~K.~M.; Gezelter,~J.~D. Simulations of Heat Conduction at
  Thiolate-Capped Gold Surfaces: The Role of Chain Length and Solvent
  Penetration. \emph{J. Phys. Chem. C} \textbf{2013}, \emph{117},
  7605--7612\relax
\mciteBstWouldAddEndPuncttrue
\mciteSetBstMidEndSepPunct{\mcitedefaultmidpunct}
{\mcitedefaultendpunct}{\mcitedefaultseppunct}\relax
\EndOfBibitem
\bibitem[Stocker \latin{et~al.}(2016)Stocker, Neidhart, and Gezelter]{ks16}
Stocker,~K.~M.; Neidhart,~S.~M.; Gezelter,~J.~D. Interfacial thermal
  conductance of thiolate-protected gold nanospheres. \emph{J. Appl. Phys.}
  \textbf{2016}, \emph{119}, 025106\relax
\mciteBstWouldAddEndPuncttrue
\mciteSetBstMidEndSepPunct{\mcitedefaultmidpunct}
{\mcitedefaultendpunct}{\mcitedefaultseppunct}\relax
\EndOfBibitem
\bibitem[Kuang and Gezelter(2011)Kuang, and Gezelter]{sk11}
Kuang,~S.; Gezelter,~J. Simulating Interfacial Thermal Conductance at
  Metal-Solvent Interfaces: The Role of Chemical Capping Agents. \emph{J. Phys.
  Chem. C} \textbf{2011}, \emph{115}, 22475--22483\relax
\mciteBstWouldAddEndPuncttrue
\mciteSetBstMidEndSepPunct{\mcitedefaultmidpunct}
{\mcitedefaultendpunct}{\mcitedefaultseppunct}\relax
\EndOfBibitem
\bibitem[Neidhart and Gezelter(2020)Neidhart, and Gezelter]{smn20}
Neidhart,~S.; Gezelter,~J. Thermal Conductivity of Gold-Phenylethanethiol
  (Au144PET60) Nanoarrays: A Molecular Dynamics Study. \emph{J. Phys. Chem.}
  \textbf{2020}, \emph{124}, 3389--3395\relax
\mciteBstWouldAddEndPuncttrue
\mciteSetBstMidEndSepPunct{\mcitedefaultmidpunct}
{\mcitedefaultendpunct}{\mcitedefaultseppunct}\relax
\EndOfBibitem
\bibitem[Salassi \latin{et~al.}(2020)Salassi, Cardellini, Asinari, Ferrando,
  and Rossi]{ss20}
Salassi,~S.; Cardellini,~A.; Asinari,~P.; Ferrando,~R.; Rossi,~G. Water
  dynamics affects thermal transport at the surface of hydrophobic and
  hydrophilic irradiated nanoparticles. \emph{Nanoscale Adv.} \textbf{2020},
  \emph{2}, 3181--3190\relax
\mciteBstWouldAddEndPuncttrue
\mciteSetBstMidEndSepPunct{\mcitedefaultmidpunct}
{\mcitedefaultendpunct}{\mcitedefaultseppunct}\relax
\EndOfBibitem
\bibitem[Hung \latin{et~al.}(2016)Hung, Kikugawa, and Shiomi]{sh16}
Hung,~S.-W.; Kikugawa,~G.; Shiomi,~J. Mechanism of Temperature Dependent
  Thermal Transport across the Interface between Self-Assembled Monolayer and
  Water. \emph{J. Phys. Chem. C} \textbf{2016}, \emph{102}, 26678--26685\relax
\mciteBstWouldAddEndPuncttrue
\mciteSetBstMidEndSepPunct{\mcitedefaultmidpunct}
{\mcitedefaultendpunct}{\mcitedefaultseppunct}\relax
\EndOfBibitem
\bibitem[Tascini \latin{et~al.}(2017)Tascini, Armstrong, Chiavazzo, Fasano,
  Asinari, and Bresme]{ast17}
Tascini,~A.~S.; Armstrong,~J.; Chiavazzo,~E.; Fasano,~M.; Asinari,~P.;
  Bresme,~F. Thermal transport across nanoparticle-fluid interfaces: the
  interplay of interfacial curvature and nanoparticle-fluid interactions.
  \emph{Phys. Chem. Chem. Phys.} \textbf{2017}, \emph{19}, 3244--3253\relax
\mciteBstWouldAddEndPuncttrue
\mciteSetBstMidEndSepPunct{\mcitedefaultmidpunct}
{\mcitedefaultendpunct}{\mcitedefaultseppunct}\relax
\EndOfBibitem
\bibitem[Jiang \latin{et~al.}(2022)Jiang, Olarte-Plata, and Bresme]{mj22}
Jiang,~M.; Olarte-Plata,~J.~D.; Bresme,~F. Heterogeneous thermal conductance of
  nanoparticle-fluid interfaces: An atomistic nodal approach. \emph{J. Chem.
  Phys.} \textbf{2022}, \emph{156}, 044701\relax
\mciteBstWouldAddEndPuncttrue
\mciteSetBstMidEndSepPunct{\mcitedefaultmidpunct}
{\mcitedefaultendpunct}{\mcitedefaultseppunct}\relax
\EndOfBibitem
\bibitem[Stocker and Gezelter(2014)Stocker, and Gezelter]{kms14}
Stocker,~K.; Gezelter,~J. A Method for Creating Thermal and Angular Momentum
  Fluxes in Nonperiodic Simulations. \emph{J. Chem. Theory Comput.}
  \textbf{2014}, \emph{10}, 1878--1886\relax
\mciteBstWouldAddEndPuncttrue
\mciteSetBstMidEndSepPunct{\mcitedefaultmidpunct}
{\mcitedefaultendpunct}{\mcitedefaultseppunct}\relax
\EndOfBibitem
\bibitem[Wilson \latin{et~al.}(2022)Wilson, Nielsen, Randrianalisoa, and
  Qin]{baw22}
Wilson,~B.~A.; Nielsen,~S.~O.; Randrianalisoa,~J.~H.; Qin,~Z. Curvature and
  temperature-dependent thermal interface conductance between nanoscale gold
  and water. \emph{J. Chem. Phys.} \textbf{2022}, \emph{157}, 054703\relax
\mciteBstWouldAddEndPuncttrue
\mciteSetBstMidEndSepPunct{\mcitedefaultmidpunct}
{\mcitedefaultendpunct}{\mcitedefaultseppunct}\relax
\EndOfBibitem
\bibitem[Paniagua-Guerra and Ramos-Alvarado(2023)Paniagua-Guerra, and
  Ramos-Alvarado]{lep23}
Paniagua-Guerra,~L.~E.; Ramos-Alvarado,~B. Thermal transport across flat and
  curved gold-water interfaces: Assessing the effects of the interfacial
  modeling parameters. \emph{J. Chem. Phys.} \textbf{2023}, \emph{158},
  134717\relax
\mciteBstWouldAddEndPuncttrue
\mciteSetBstMidEndSepPunct{\mcitedefaultmidpunct}
{\mcitedefaultendpunct}{\mcitedefaultseppunct}\relax
\EndOfBibitem
\bibitem[Guti\'errez-Varela \latin{et~al.}(2022)Guti\'errez-Varela, Merabia,
  and Santamaria]{og22}
Guti\'errez-Varela,~O.; Merabia,~S.; Santamaria,~R. Size-dependent effects of
  the thermal transport at gold nanoparticle-water interfaces. \emph{J. Chem.
  Phys.} \textbf{2022}, \emph{157}, 084702\relax
\mciteBstWouldAddEndPuncttrue
\mciteSetBstMidEndSepPunct{\mcitedefaultmidpunct}
{\mcitedefaultendpunct}{\mcitedefaultseppunct}\relax
\EndOfBibitem
\bibitem[Pandey and Leitner(2016)Pandey, and Leitner]{hdp16}
Pandey,~H.~D.; Leitner,~D.~M. Thermalization and Thermal Transport in
  Molecules. \emph{J. Phys. Chem. Lett.} \textbf{2016}, \emph{7},
  5062--5067\relax
\mciteBstWouldAddEndPuncttrue
\mciteSetBstMidEndSepPunct{\mcitedefaultmidpunct}
{\mcitedefaultendpunct}{\mcitedefaultseppunct}\relax
\EndOfBibitem
\bibitem[Bhattarai \latin{et~al.}(2019)Bhattarai, Newman, and Gezelter]{hb19}
Bhattarai,~H.; Newman,~K.; Gezelter,~J. Polarizable potentials for metals: The
  density readjusting embedded atom method (DR-EAM). \emph{Phys. Rev. B}
  \textbf{2019}, \emph{99}, 094106\relax
\mciteBstWouldAddEndPuncttrue
\mciteSetBstMidEndSepPunct{\mcitedefaultmidpunct}
{\mcitedefaultendpunct}{\mcitedefaultseppunct}\relax
\EndOfBibitem
\bibitem[Bhattarai \latin{et~al.}(2020)Bhattarai, Newman, and Gezelter]{hb20}
Bhattarai,~H.; Newman,~K.; Gezelter,~J. The role of polarizability in the
  interfacial thermal conductance at the gold-water interface. \emph{J. Chem.
  Phys.} \textbf{2020}, \emph{153}, 204703\relax
\mciteBstWouldAddEndPuncttrue
\mciteSetBstMidEndSepPunct{\mcitedefaultmidpunct}
{\mcitedefaultendpunct}{\mcitedefaultseppunct}\relax
\EndOfBibitem
\bibitem[Shavalier and Gezelter(2022)Shavalier, and Gezelter]{ss22}
Shavalier,~S.~A.; Gezelter,~J.~D. Thermal Transport in Citrate-Capped
  Interfaces Using a Polarizable Force Field. \emph{J. Phys. Chem. C}
  \textbf{2022}, \emph{126}, 12742--12754\relax
\mciteBstWouldAddEndPuncttrue
\mciteSetBstMidEndSepPunct{\mcitedefaultmidpunct}
{\mcitedefaultendpunct}{\mcitedefaultseppunct}\relax
\EndOfBibitem
\bibitem[Kuang and Gezelter(2012)Kuang, and Gezelter]{sk12}
Kuang,~S.; Gezelter,~J. Velocity shearing and scaling RNEMD: a minimally
  perturbing method for simulating temperature and momentum gradients.
  \emph{Mol. Phys.} \textbf{2012}, \emph{110:9-10}, 691--701\relax
\mciteBstWouldAddEndPuncttrue
\mciteSetBstMidEndSepPunct{\mcitedefaultmidpunct}
{\mcitedefaultendpunct}{\mcitedefaultseppunct}\relax
\EndOfBibitem
\bibitem[Zhou \latin{et~al.}(2004)Zhou, Johnson, and Wadley]{xwz04}
Zhou,~X.; Johnson,~R.; Wadley,~H. Misfit-energy-increasing dislocations in
  vapor-deposited CoFe/NiFe multilayers. \emph{Phys. Rev. B} \textbf{2004},
  \emph{69}, 144113\relax
\mciteBstWouldAddEndPuncttrue
\mciteSetBstMidEndSepPunct{\mcitedefaultmidpunct}
{\mcitedefaultendpunct}{\mcitedefaultseppunct}\relax
\EndOfBibitem
\bibitem[Berendsen \latin{et~al.}(1987)Berendsen, Grigera, and
  Straatsma]{hjcb87}
Berendsen,~H.; Grigera,~J.; Straatsma,~T. The Missing Term in Effective Pair
  Potentials. \emph{J. Phys. Chem.} \textbf{1987}, \emph{91}, 6269--6271\relax
\mciteBstWouldAddEndPuncttrue
\mciteSetBstMidEndSepPunct{\mcitedefaultmidpunct}
{\mcitedefaultendpunct}{\mcitedefaultseppunct}\relax
\EndOfBibitem
\bibitem[Lubna \latin{et~al.}(2005)Lubna, Kamath, Potoff, Rai, and
  Siepmann]{nl05}
Lubna,~N.; Kamath,~G.; Potoff,~J.~J.; Rai,~N.; Siepmann,~J.~I. Transferable
  Potentials for Phase Equilibria. 8. United-Atom Description for Thiols,
  Sulfides, Disulfides, and Thiophene. \emph{J. Phys. Chem. B} \textbf{2005},
  \emph{109}, 24100--24107\relax
\mciteBstWouldAddEndPuncttrue
\mciteSetBstMidEndSepPunct{\mcitedefaultmidpunct}
{\mcitedefaultendpunct}{\mcitedefaultseppunct}\relax
\EndOfBibitem
\bibitem[Chen \latin{et~al.}(2001)Chen, Potoff, and Siepmann]{bc01}
Chen,~B.; Potoff,~J.~J.; Siepmann,~J.~I. Monte Carlo Calculations for Alcohols
  and Their Mixtures with Alkanes. Transferable Potentials for Phase
  Equilibria. 5. United-Atom Description of Primary, Secondary, and Tertiary
  Alcohols. \emph{J. Phys. Chem. B} \textbf{2001}, \emph{105}, 3093--3104\relax
\mciteBstWouldAddEndPuncttrue
\mciteSetBstMidEndSepPunct{\mcitedefaultmidpunct}
{\mcitedefaultendpunct}{\mcitedefaultseppunct}\relax
\EndOfBibitem
\bibitem[Stubbs \latin{et~al.}(2004)Stubbs, Potoff, and Siepmann]{js04}
Stubbs,~J.~M.; Potoff,~J.~J.; Siepmann,~J.~I. Transferable Potentials for Phase
  Equilibria. 6. United-Atom Description for Ethers, Glycols, Ketones, and
  Aldehydes. \emph{J. Phys. Chem. B} \textbf{2004}, \emph{108},
  17596--17605\relax
\mciteBstWouldAddEndPuncttrue
\mciteSetBstMidEndSepPunct{\mcitedefaultmidpunct}
{\mcitedefaultendpunct}{\mcitedefaultseppunct}\relax
\EndOfBibitem
\bibitem[Weiner \latin{et~al.}(1986)Weiner, Kollman, Nguyen, and Case]{sjw86}
Weiner,~S.~J.; Kollman,~P.~A.; Nguyen,~D.~T.; Case,~D.~A. An all atom force
  field for simulations of proteins and nucleic acids. \emph{J. Comp. Chem.}
  \textbf{1986}, \emph{7}, 230--252\relax
\mciteBstWouldAddEndPuncttrue
\mciteSetBstMidEndSepPunct{\mcitedefaultmidpunct}
{\mcitedefaultendpunct}{\mcitedefaultseppunct}\relax
\EndOfBibitem
\bibitem[Jorgensen \latin{et~al.}(1996)Jorgensen, Maxwell, and
  Tirado-Rives]{wlj96}
Jorgensen,~W.~L.; Maxwell,~D.~S.; Tirado-Rives,~J. Development and Testing of
  the OPLS All-Atom Force Field on Conformational Energetics and Properties of
  Organic Liquids. \emph{J. Am. Chem. Soc.} \textbf{1996}, \emph{118},
  11225--11236\relax
\mciteBstWouldAddEndPuncttrue
\mciteSetBstMidEndSepPunct{\mcitedefaultmidpunct}
{\mcitedefaultendpunct}{\mcitedefaultseppunct}\relax
\EndOfBibitem
\bibitem[Schapotschnikow \latin{et~al.}(2007)Schapotschnikow, Pool, and
  Vlugt]{ps07}
Schapotschnikow,~P.; Pool,~R.; Vlugt,~T. J.~H. Selective adsorption of alkyl
  thiols on gold in different geometries. \emph{Comp. Phys. Comm.}
  \textbf{2007}, \emph{177}, 154--157\relax
\mciteBstWouldAddEndPuncttrue
\mciteSetBstMidEndSepPunct{\mcitedefaultmidpunct}
{\mcitedefaultendpunct}{\mcitedefaultseppunct}\relax
\EndOfBibitem
\bibitem[Dou \latin{et~al.}(2001)Dou, Zhigilei, Winograd, and Garrison]{yd01}
Dou,~Y.; Zhigilei,~L.~V.; Winograd,~N.; Garrison,~B.~J. Explosive Boiling of
  Water Films Adjacent to Heated Surfaces: A Microscopic Description. \emph{J.
  Phys. Chem. A} \textbf{2001}, \emph{105}, 2748--2755\relax
\mciteBstWouldAddEndPuncttrue
\mciteSetBstMidEndSepPunct{\mcitedefaultmidpunct}
{\mcitedefaultendpunct}{\mcitedefaultseppunct}\relax
\EndOfBibitem
\bibitem[Mart\'{i}nez \latin{et~al.}(2009)Mart\'{i}nez, Andrade, Birgin, and
  Mart\'{i}nez]{pack}
Mart\'{i}nez,~L.; Andrade,~R.; Birgin,~E.; Mart\'{i}nez,~J. Packmol: A package
  for building initial configurations for molecular dynamics simulations.
  \emph{J. Chem. Comput.} \textbf{2009}, \emph{30(13)}, 2157--2164\relax
\mciteBstWouldAddEndPuncttrue
\mciteSetBstMidEndSepPunct{\mcitedefaultmidpunct}
{\mcitedefaultendpunct}{\mcitedefaultseppunct}\relax
\EndOfBibitem
\bibitem[Hoover(1985)]{wgh85}
Hoover,~W.~G. Canonical dynamics: Equilibrium phase-space distributions.
  \emph{Phys. Rev. A} \textbf{1985}, \emph{31}, 1695--1697\relax
\mciteBstWouldAddEndPuncttrue
\mciteSetBstMidEndSepPunct{\mcitedefaultmidpunct}
{\mcitedefaultendpunct}{\mcitedefaultseppunct}\relax
\EndOfBibitem
\bibitem[Melchionna \latin{et~al.}(1993)Melchionna, Ciccotti, and Holian]{sm93}
Melchionna,~S.; Ciccotti,~G.; Holian,~B.~L. Hoover NPT dynamics for systems
  varying in shape and size. \emph{Mol. Phys.} \textbf{1993}, \emph{78},
  533--544\relax
\mciteBstWouldAddEndPuncttrue
\mciteSetBstMidEndSepPunct{\mcitedefaultmidpunct}
{\mcitedefaultendpunct}{\mcitedefaultseppunct}\relax
\EndOfBibitem
\bibitem[Vardeman \latin{et~al.}(2011)Vardeman, Stocker, and Gezelter]{cfv11}
Vardeman,~C.~F.; Stocker,~K.~M.; Gezelter,~J.~D. The Langevin Hull: Constant
  Pressure and Temperature Dynamics for Nonperiodic Systems. \emph{J. Chem.
  Theory Comput.} \textbf{2011}, \emph{7}, 834--842\relax
\mciteBstWouldAddEndPuncttrue
\mciteSetBstMidEndSepPunct{\mcitedefaultmidpunct}
{\mcitedefaultendpunct}{\mcitedefaultseppunct}\relax
\EndOfBibitem
\bibitem[Gezelter \latin{et~al.}()Gezelter, Bhattarai, Drisko, Duraes, Lin,
  Vardeman, Fennell, Meineke, Louden, Neidhart, Kuang, Lamichhane, Michalka,
  Stocker, Marr, Sun, Li, Daily, and Zheng]{openmd}
Gezelter,~J.; Bhattarai,~H.; Drisko,~C.; Duraes,~A.; Lin,~T.; Vardeman,~C.;
  Fennell,~C.; Meineke,~M.; Louden,~P.; Neidhart,~S. \latin{et~al.}
  \emph{OpenMD, an Open Source Engine for Molecular Dynamics}, Version 2.7;
  http://openmd.org. Accessed December 2, 2022. \relax
\mciteBstWouldAddEndPunctfalse
\mciteSetBstMidEndSepPunct{\mcitedefaultmidpunct}
{}{\mcitedefaultseppunct}\relax
\EndOfBibitem
\bibitem[Swartz and Pohl(1989)Swartz, and Pohl]{ets89}
Swartz,~E.; Pohl,~R. Thermal boundary resistance. \emph{Rev. Mod. Phys.}
  \textbf{1989}, \emph{61}, 605--668\relax
\mciteBstWouldAddEndPuncttrue
\mciteSetBstMidEndSepPunct{\mcitedefaultmidpunct}
{\mcitedefaultendpunct}{\mcitedefaultseppunct}\relax
\EndOfBibitem
\bibitem[Monachon \latin{et~al.}(2016)Monachon, Weber, and Dames]{cm16}
Monachon,~C.; Weber,~L.; Dames,~C. Thermal Boundary Conductance: A Materials
  Science Perspective. \emph{Annu. Rev. Mater. Res.} \textbf{2016}, \emph{46},
  433--463\relax
\mciteBstWouldAddEndPuncttrue
\mciteSetBstMidEndSepPunct{\mcitedefaultmidpunct}
{\mcitedefaultendpunct}{\mcitedefaultseppunct}\relax
\EndOfBibitem
\end{mcitethebibliography}


\providecommand{\latin}[1]{#1}
\makeatletter
\providecommand{\doi}
  {\begingroup\let\do\@makeother\dospecials
  \catcode`\{=1 \catcode`\}=2 \doi@aux}
\providecommand{\doi@aux}[1]{\endgroup\texttt{#1}}
\makeatother
\providecommand*\mcitethebibliography{\thebibliography}
\csname @ifundefined\endcsname{endmcitethebibliography}
  {\let\endmcitethebibliography\endthebibliography}{}
\begin{mcitethebibliography}{21}
\providecommand*\natexlab[1]{#1}
\providecommand*\mciteSetBstSublistMode[1]{}
\providecommand*\mciteSetBstMaxWidthForm[2]{}
\providecommand*\mciteBstWouldAddEndPuncttrue
  {\def\EndOfBibitem{\unskip.}}
\providecommand*\mciteBstWouldAddEndPunctfalse
  {\let\EndOfBibitem\relax}
\providecommand*\mciteSetBstMidEndSepPunct[3]{}
\providecommand*\mciteSetBstSublistLabelBeginEnd[3]{}
\providecommand*\EndOfBibitem{}
\mciteSetBstSublistMode{f}
\mciteSetBstMaxWidthForm{subitem}{(\alph{mcitesubitemcount})}
\mciteSetBstSublistLabelBeginEnd
  {\mcitemaxwidthsubitemform\space}
  {\relax}
  {\relax}

\bibitem[Stubbs \latin{et~al.}(2004)Stubbs, Potoff, and Siepmann]{js04}
Stubbs,~J.~M.; Potoff,~J.~J.; Siepmann,~J.~I. Transferable Potentials for Phase
  Equilibria. 6. United-Atom Description for Ethers, Glycols, Ketones, and
  Aldehydes. \emph{J. Phys. Chem. B} \textbf{2004}, \emph{108},
  17596--17605\relax
\mciteBstWouldAddEndPuncttrue
\mciteSetBstMidEndSepPunct{\mcitedefaultmidpunct}
{\mcitedefaultendpunct}{\mcitedefaultseppunct}\relax
\EndOfBibitem
\bibitem[Lubna \latin{et~al.}(2005)Lubna, Kamath, Potoff, Rai, and
  Siepmann]{nl05}
Lubna,~N.; Kamath,~G.; Potoff,~J.~J.; Rai,~N.; Siepmann,~J.~I. Transferable
  Potentials for Phase Equilibria. 8. United-Atom Description for Thiols,
  Sulfides, Disulfides, and Thiophene. \emph{J. Phys. Chem. B} \textbf{2005},
  \emph{109}, 24100--24107\relax
\mciteBstWouldAddEndPuncttrue
\mciteSetBstMidEndSepPunct{\mcitedefaultmidpunct}
{\mcitedefaultendpunct}{\mcitedefaultseppunct}\relax
\EndOfBibitem
\bibitem[Schapotschnikow \latin{et~al.}(2007)Schapotschnikow, Pool, and
  Vlugt]{ps07}
Schapotschnikow,~P.; Pool,~R.; Vlugt,~T. J.~H. Selective adsorption of alkyl
  thiols on gold in different geometries. \emph{Comp. Phys. Comm.}
  \textbf{2007}, \emph{177}, 154--157\relax
\mciteBstWouldAddEndPuncttrue
\mciteSetBstMidEndSepPunct{\mcitedefaultmidpunct}
{\mcitedefaultendpunct}{\mcitedefaultseppunct}\relax
\EndOfBibitem
\bibitem[Weiner \latin{et~al.}(1986)Weiner, Kollman, Nguyen, and Case]{sjw86}
Weiner,~S.~J.; Kollman,~P.~A.; Nguyen,~D.~T.; Case,~D.~A. An all atom force
  field for simulations of proteins and nucleic acids. \emph{J. Comp. Chem.}
  \textbf{1986}, \emph{7}, 230--252\relax
\mciteBstWouldAddEndPuncttrue
\mciteSetBstMidEndSepPunct{\mcitedefaultmidpunct}
{\mcitedefaultendpunct}{\mcitedefaultseppunct}\relax
\EndOfBibitem
\bibitem[Martin and Siepmann(1998)Martin, and Siepmann]{mgm98}
Martin,~M.~G.; Siepmann,~J.~I. Transferable Potentials for Phase Equilibria. 1.
  United-Atom Description of n-Alkanes. \emph{J. Phys. Chem. B} \textbf{1998},
  \emph{102}, 2569--2577\relax
\mciteBstWouldAddEndPuncttrue
\mciteSetBstMidEndSepPunct{\mcitedefaultmidpunct}
{\mcitedefaultendpunct}{\mcitedefaultseppunct}\relax
\EndOfBibitem
\bibitem[Chen \latin{et~al.}(2001)Chen, Potoff, and Siepmann]{bc01}
Chen,~B.; Potoff,~J.~J.; Siepmann,~J.~I. Monte Carlo Calculations for Alcohols
  and Their Mixtures with Alkanes. Transferable Potentials for Phase
  Equilibria. 5. United-Atom Description of Primary, Secondary, and Tertiary
  Alcohols. \emph{J. Phys. Chem. B} \textbf{2001}, \emph{105}, 3093--3104\relax
\mciteBstWouldAddEndPuncttrue
\mciteSetBstMidEndSepPunct{\mcitedefaultmidpunct}
{\mcitedefaultendpunct}{\mcitedefaultseppunct}\relax
\EndOfBibitem
\bibitem[Shavalier and Gezelter(2022)Shavalier, and Gezelter]{ss22}
Shavalier,~S.~A.; Gezelter,~J.~D. Thermal Transport in Citrate-Capped
  Interfaces Using a Polarizable Force Field. \emph{J. Phys. Chem. C}
  \textbf{2022}, \emph{126}, 12742--12754\relax
\mciteBstWouldAddEndPuncttrue
\mciteSetBstMidEndSepPunct{\mcitedefaultmidpunct}
{\mcitedefaultendpunct}{\mcitedefaultseppunct}\relax
\EndOfBibitem
\bibitem[Berg \latin{et~al.}(2017)Berg, Peter, and Johnston]{ab17}
Berg,~A.; Peter,~C.; Johnston,~K. Evaluation and Optimisation of Interface
  Force Fields for Water on Gold Surfaces. \emph{J. Chem. Theory Comput.}
  \textbf{2017}, \emph{13}, 5610--5623\relax
\mciteBstWouldAddEndPuncttrue
\mciteSetBstMidEndSepPunct{\mcitedefaultmidpunct}
{\mcitedefaultendpunct}{\mcitedefaultseppunct}\relax
\EndOfBibitem
\bibitem[Zhou \latin{et~al.}(2004)Zhou, Johnson, and Wadley]{xwz04}
Zhou,~X.; Johnson,~R.; Wadley,~H. Misfit-energy-increasing dislocations in
  vapor-deposited CoFe/NiFe multilayers. \emph{Phys. Rev. B} \textbf{2004},
  \emph{69}, 144113\relax
\mciteBstWouldAddEndPuncttrue
\mciteSetBstMidEndSepPunct{\mcitedefaultmidpunct}
{\mcitedefaultendpunct}{\mcitedefaultseppunct}\relax
\EndOfBibitem
\bibitem[Bhattarai \latin{et~al.}(2019)Bhattarai, Newman, and Gezelter]{hb19}
Bhattarai,~H.; Newman,~K.; Gezelter,~J. Polarizable potentials for metals: The
  density readjusting embedded atom method (DR-EAM). \emph{Phys. Rev. B}
  \textbf{2019}, \emph{99}, 094106\relax
\mciteBstWouldAddEndPuncttrue
\mciteSetBstMidEndSepPunct{\mcitedefaultmidpunct}
{\mcitedefaultendpunct}{\mcitedefaultseppunct}\relax
\EndOfBibitem
\bibitem[Fennell and Gezelter(2006)Fennell, and Gezelter]{cjf06}
Fennell,~C.; Gezelter,~J. Is the Ewald summation still necessary? Pairwise
  alternatives to the accepted standard for long-range electrostatics. \emph{J.
  Chem. Phys.} \textbf{2006}, \emph{124}, 234104\relax
\mciteBstWouldAddEndPuncttrue
\mciteSetBstMidEndSepPunct{\mcitedefaultmidpunct}
{\mcitedefaultendpunct}{\mcitedefaultseppunct}\relax
\EndOfBibitem
\bibitem[Dou \latin{et~al.}(2001)Dou, Zhigilei, Winograd, and Garrison]{yd01}
Dou,~Y.; Zhigilei,~L.~V.; Winograd,~N.; Garrison,~B.~J. Explosive Boiling of
  Water Films Adjacent to Heated Surfaces: A Microscopic Description. \emph{J.
  Phys. Chem. A} \textbf{2001}, \emph{105}, 2748--2755\relax
\mciteBstWouldAddEndPuncttrue
\mciteSetBstMidEndSepPunct{\mcitedefaultmidpunct}
{\mcitedefaultendpunct}{\mcitedefaultseppunct}\relax
\EndOfBibitem
\bibitem[Pool \latin{et~al.}(2007)Pool, Schapotschnikow, and Vlugt]{rp07}
Pool,~R.; Schapotschnikow,~P.; Vlugt,~T. J.~H. Solvent Effects in the
  Adsorption of Alkyl Thiols on Gold Structures: A Molecular Simulation Study.
  \emph{J. Phys. Chem. C} \textbf{2007}, \emph{111}, 10201--10212\relax
\mciteBstWouldAddEndPuncttrue
\mciteSetBstMidEndSepPunct{\mcitedefaultmidpunct}
{\mcitedefaultendpunct}{\mcitedefaultseppunct}\relax
\EndOfBibitem
\bibitem[Berendsen \latin{et~al.}(1987)Berendsen, Grigera, and
  Straatsma]{hjcb87}
Berendsen,~H.; Grigera,~J.; Straatsma,~T. The Missing Term in Effective Pair
  Potentials. \emph{J. Phys. Chem.} \textbf{1987}, \emph{91}, 6269--6271\relax
\mciteBstWouldAddEndPuncttrue
\mciteSetBstMidEndSepPunct{\mcitedefaultmidpunct}
{\mcitedefaultendpunct}{\mcitedefaultseppunct}\relax
\EndOfBibitem
\bibitem[Bhattarai \latin{et~al.}(2020)Bhattarai, Newman, and Gezelter]{hb20}
Bhattarai,~H.; Newman,~K.; Gezelter,~J. The role of polarizability in the
  interfacial thermal conductance at the gold-water interface. \emph{J. Chem.
  Phys.} \textbf{2020}, \emph{153}, 204703\relax
\mciteBstWouldAddEndPuncttrue
\mciteSetBstMidEndSepPunct{\mcitedefaultmidpunct}
{\mcitedefaultendpunct}{\mcitedefaultseppunct}\relax
\EndOfBibitem
\bibitem[Jorgensen \latin{et~al.}(1996)Jorgensen, Maxwell, and
  Tirado-Rives]{wlj96}
Jorgensen,~W.~L.; Maxwell,~D.~S.; Tirado-Rives,~J. Development and Testing of
  the OPLS All-Atom Force Field on Conformational Energetics and Properties of
  Organic Liquids. \emph{J. Am. Chem. Soc.} \textbf{1996}, \emph{118},
  11225--11236\relax
\mciteBstWouldAddEndPuncttrue
\mciteSetBstMidEndSepPunct{\mcitedefaultmidpunct}
{\mcitedefaultendpunct}{\mcitedefaultseppunct}\relax
\EndOfBibitem
\bibitem[Gezelter \latin{et~al.}()Gezelter, Bhattarai, Drisko, Duraes, Lin,
  Vardeman, Fennell, Meineke, Louden, Neidhart, Kuang, Lamichhane, Michalka,
  Stocker, Marr, Sun, Li, Daily, and Zheng]{openmd}
Gezelter,~J.; Bhattarai,~H.; Drisko,~C.; Duraes,~A.; Lin,~T.; Vardeman,~C.;
  Fennell,~C.; Meineke,~M.; Louden,~P.; Neidhart,~S.; Kuang,~S.;
  Lamichhane,~M.; Michalka,~J.; Stocker,~K.; Marr,~J.; Sun,~X.; Li,~C.;
  Daily,~K.; Zheng,~Y. \emph{OpenMD, an Open Source Engine for Molecular
  Dynamics}, Version 2.7; http://openmd.org. Accessed December 2, 2022. \relax
\mciteBstWouldAddEndPunctfalse
\mciteSetBstMidEndSepPunct{\mcitedefaultmidpunct}
{}{\mcitedefaultseppunct}\relax
\EndOfBibitem
\bibitem[Mart\'{i}nez \latin{et~al.}(2009)Mart\'{i}nez, Andrade, Birgin, and
  Mart\'{i}nez]{pack}
Mart\'{i}nez,~L.; Andrade,~R.; Birgin,~E.; Mart\'{i}nez,~J. Packmol: A package
  for building initial configurations for molecular dynamics simulations.
  \emph{J. Chem. Comput.} \textbf{2009}, \emph{30(13)}, 2157--2164\relax
\mciteBstWouldAddEndPuncttrue
\mciteSetBstMidEndSepPunct{\mcitedefaultmidpunct}
{\mcitedefaultendpunct}{\mcitedefaultseppunct}\relax
\EndOfBibitem
\bibitem[Vardeman \latin{et~al.}(2011)Vardeman, Stocker, and Gezelter]{cfv11}
Vardeman,~C.~F.; Stocker,~K.~M.; Gezelter,~J.~D. The Langevin Hull: Constant
  Pressure and Temperature Dynamics for Nonperiodic Systems. \emph{J. Chem.
  Theory Comput.} \textbf{2011}, \emph{7}, 834--842\relax
\mciteBstWouldAddEndPuncttrue
\mciteSetBstMidEndSepPunct{\mcitedefaultmidpunct}
{\mcitedefaultendpunct}{\mcitedefaultseppunct}\relax
\EndOfBibitem
\bibitem[Stocker and Gezelter(2013)Stocker, and Gezelter]{ks13}
Stocker,~K.~M.; Gezelter,~J.~D. Simulations of Heat Conduction at
  Thiolate-Capped Gold Surfaces: The Role of Chain Length and Solvent
  Penetration. \emph{J. Phys. Chem. C} \textbf{2013}, \emph{117},
  7605--7612\relax
\mciteBstWouldAddEndPuncttrue
\mciteSetBstMidEndSepPunct{\mcitedefaultmidpunct}
{\mcitedefaultendpunct}{\mcitedefaultseppunct}\relax
\EndOfBibitem
\end{mcitethebibliography}

\end{document}

\begin{abstract}
This supporting document provides details of the force field parameters, information on simulation construction and composition,  method validation,  and equilibration details for systems described in the accompanying paper. We also include  vibrational power spectra for all system components, Bhattacharyya coefficients computed from vibrational power spectra,  thermal profiles, mass densities,  and charge densities near the interfaces of all polarizable systems.
\end{abstract}

\section{System Details}
This section contains parameters for non-bonded interactions (Table \ref{tab:nb}), a soft sphere repulsive interaction (Table \ref{tab:rep}), harmonic bonds (Table \ref{tab:bond}), harmonic bends (Table \ref{tab:bend}), torsions (Table \ref{tab:torsion}), embedded atom method (EAM) parameters (Tables \ref{tab:eam_phi} and \ref{tab:eam_rho}), and density-readjusting embedded atom method (DR-EAM) parameters (Table \ref{tab:dream}).

\subsection{Force Fields}

\begin{figure}[h]
    \centering
    \includegraphics[width=\linewidth]{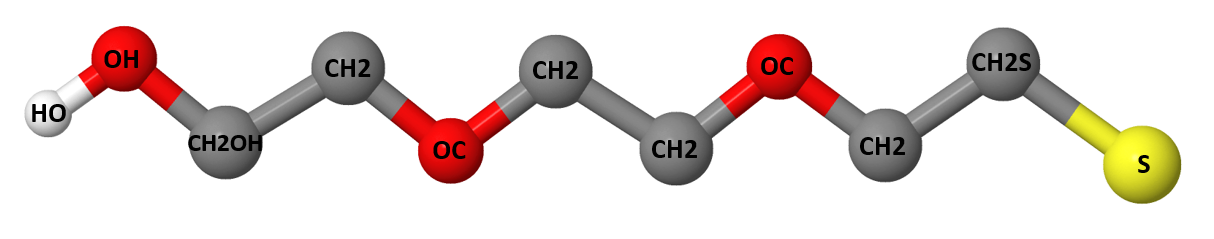}
    \caption{thiolated PEG atom types. Carbon atoms are gray, oxygen atoms are red, sulfur is yellow, and the explicit hydrogen is white. All other hydrogen atoms are treated implicitly.\label{fig:thiolpeg}} 
\end{figure}

The thiolated polyethylene glycol (PEG) interaction parameters are taken from previous force fields describing ethers\cite{js04} and various thiols.\cite{nl05, ps07}. We use nonbonded interactions,\cite{js04, nl05, ps07} harmonic bonds,\cite{js04, sjw86, mgm98, bc01} harmonic bends,\cite{js04, nl05, mgm98, bc01} and both TraPPE\cite{js04, nl05, mgm98, bc01} and CHARMM\cite{js04, mgm98, bc01} torsions to describe the thiolated PEG molecules. The atom types for thiolated PEG are shown in Figure \ref{fig:thiolpeg}. To prevent unphysical collisions between the hydroxyl hydrogen (HO) and ether oxygens (OC), an additional soft sphere $(r^{-12})$ repulsion is added between these two atom types.

Rigid SPC/E water is used as a solvent in this work. The interactions between water and gold are identical to our previous work.\cite{ss22}  These interaction parameters yield an $E_\text{{abs}}$ = -19.5 kJ/mol for water on a (111) gold surface, which is comparable to DFT calculations.\cite{ab17}

As pair potentials alone are not sufficient to describe the many-atom properties of metals, we use EAM parameters from Zhou \textit{et al.} for gold in non-polarizable systems.\cite{xwz04}. We use DR-EAM parameters from Bhattarai \textit{et al.} for gold in polarizable systems.\cite{hb19}. The damped shifted force model from Fennell and Gezelter was used for long-range electrostatic interactions with a damping parameter, $\alpha = 0.18~\mathrm{\AA}^{-1}$, which yields good agreement with Ewald electrostatics when $r_\mathrm{cut} = 12~\mathrm{\AA}$.\cite{cjf06}

\begin{sidewaystable}[h]               
\begin{center}
\caption{Non-bonded interactions used for simulations of gold nanostructures and thiolated PEG in a water solvent. All parameters listed are Lennard-Jones $n-m$ pairwise terms. The $\epsilon_{\text{Au}-i}$ values for water have been modified from Dou \textit{et al.}\cite{yd01} to more closely replicate the adsorption energy for water on a (111) gold surface determined by DFT calculations. \cite{ab17}. The cross interactions containing HO and H-SPCE are purely repulsive.} \label{tab:nb}  
\bibpunct{}{}{,}{n}{,}{,}
\begin{tabular}{ |c c c c c c c c c c| }      
\toprule                 
\text{Atom} & \text{Mass (au)} & $\sigma_{ii}$ (\AA{}) & $\epsilon_{ii}$ (kcal/mol) & $\sigma_{\text{Au}-i}$ (\AA{}) & $\epsilon_{\text{Au}-i}$ (kcal/mol) & n & m & Charge (e) & Source \\
\hline
\ce{Au} & 196.97 & - & - & - & - & - & - & 0.0 &\\
\ce{\text{CH2}} & 14.03 &  3.95$^{a}$ &  0.0914$^{a}$ & 3.54$^{b}$ & 0.175$^{b}$ & 12 & 6 & 0.250$^{a}$ & Refs. \protect\cite{js04}$^{a}$, \protect\cite{ps07}$^{b}$\\
\ce{\text{CH2OH}} & 14.03 & 3.95$^{a}$ & 0.0914$^{a}$ & 3.54$^{b}$ & 0.175$^{b}$ & 12 & 6 & 0.265$^{a}$ & Refs. \protect\cite{js04}$^{a}$, \protect\cite{ps07}$^{b}$\\
\ce{\text{CH2S}} & 14.03 & 3.95$^{a}$ & 0.0914$^{a}$ & 3.54$^{b}$ & 0.175$^{b}$ & 12 & 6 & 0.171$^{a}$ & Refs. \protect\cite{nl05}$^{a}$, \protect\cite{ps07}$^{b}$\\
\ce{\text{OH}} &  15.9994 & 3.02$^{a}$ & 0.1848$^{a}$ & 3.12$^{a,b}$ & 0.213$^{a,b}$ & 12 & 6 & -0.700$^{a}$ & Refs. \protect\cite{js04}$^{a}$, \protect\cite{ps07}$^{b}$\\
\ce{\text{HO}} & 1.0079 &  0.0 & 0.0 & 1 & 79487.7$^{a}$ & 12 & - & 0.435$^{a}$ & Ref. \protect\cite{js04}\\
\ce{\text{OC}} & 15.9994 &  2.80$^{a}$ & 0.1093$^{a}$ & 3.01$^{a,b}$ & 0.164$^{a,b}$ & 12 & 6 & -0.500$^{a}$ & Refs. \protect\cite{js04}$^{a}$, \protect\cite{ps07}$^{b}$\\
\ce{\text{S}} & 32.0655 & 3.62$^{a}$ & 0.4610$^{a}$ & 2.65$^{b}$ & 5.554$^{b}$ & 12 & 6 & -0.171$^{a}$ & Refs. \protect\cite{nl05}$^{a}$, \protect\cite{rp07}$^{b}$\\
\ce{O-SPCE} & 15.9994 & 3.16549$^{a}$ & 0.15532$^{a}$ & 3.18284 & 0.6285748 & 9 & 6 & -0.848 & Refs. \protect\cite{hjcb87}, \protect\cite{hb20}\\
\ce{H-SPCE} &  1.0079 & 0.0 & 0.0 & 1 & 0.0008492 & 4 & - & 0.424 & Refs. \protect\cite{hjcb87}, \protect\cite{hb20}\\
\hline
\end{tabular}
\end{center}
\end{sidewaystable}

\clearpage

\begin{table}[h]                
\bibpunct{}{}{,}{n}{,}{,}
\begin{center}
\caption{Repulsive soft sphere $(r^{-12})$ interaction between the hydroxyl hydrogen and non-terminal ether oxygen (OC) atoms in the thiolated PEG.}
\label{tab:rep}
\begin{tabular}{ |c c c c c c| }      
\hline                 
\emph{$i$} & \emph{$j$} & $\sigma_{ij}$ (\AA{}) & $\epsilon_{ij}$ (kcal/mol) & n & Source\\
\hline
\ce{\text{HO}} & \text{OC} & 1 & 79487.7 & 12 & Ref. \protect\cite{js04}\\
\hline
\end{tabular}
\end{center}
\end{table}

\begin{table}[h]                
\bibpunct{}{}{,}{n}{,}{,}
\begin{center}
\caption{Harmonic bond parameters for thiolated PEG. Water is simulated as a rigid body so no harmonic bond parameters are needed.}
\label{tab:bond}
\begin{tabular}{ |c c c c c| }      
\hline                 
\emph{$i$} & \emph{$j$} & $r_{0}$ (\AA{}) & $K_{ij}$ (kcal/mol $\times$ \text{\AA{}}$^2$) & Source\\
\hline
\ce{\text{CH2}} & \text{OC} & 1.41$^{a}$ & 640.0$^{b}$ & Refs. \protect\cite{js04}$^{a}$, \protect\cite{sjw86}$^{b}$\\
\ce{\text{CH2}} & \text{CH2} & 1.54$^{a}$ & 536.0$^{b}$ & Refs. \protect\cite{js04}$^{a}$, \protect\cite{wlj96}$^{b}$\\
\ce{\text{CH2}} & \text{CH2S} & 1.54$^{a}$ & 536.0$^{b}$ & Refs. \protect\cite{js04}$^{a}$, \protect\cite{wlj96}$^{b}$\\
\ce{\text{OH}} & \text{HO} & 0.945$^{a}$ & 553.0$^{b}$ & Refs. \protect\cite{bc01}$^{a}$, \protect\cite{sjw86}$^{b}$\\
\ce{\text{S}} & \text{CH2S} & 1.81 & 444.0 & Ref. \protect\cite{sjw86}\\
\ce{\text{CH2OH}} & \text{CH2} & 1.54$^{a}$ & 536.0$^{b}$ & Refs. \protect\cite{js04}$^{a}$, \protect\cite{wlj96}$^{b}$\\
\ce{\text{OH}} & \text{CH2OH} & 1.41$^{a}$ & 640.0$^{b}$ & Refs. \protect\cite{js04}$^{a}$, \protect\cite{sjw86}$^{b}$\\
\hline
\end{tabular}
\end{center}
\end{table}

\begin{table}[h]              
\bibpunct{}{}{,}{n}{,}{,}
\begin{center}
\caption{Harmonic bend parameters for thiolated PEG. The central atom in each bend is atom $j$, bonded to atoms $i$ and $k$.}
\label{tab:bend}    
\begin{tabular}{ |c c c c c c| }     
\hline                 
\emph{i} & \emph{j} & \emph{k} & $\theta_0$ (degrees) & $K_{ijk}$ (kcal/mol $\times$ \text{rad}$^2$) & Source\\
\hline
\ce{\text{HO}} & \text{OH} & \text{CH2OH} & 108.5 & 110.0844 & Ref. \protect\cite{bc01}\\
\ce{\text{CH2}} & \text{CH2} & \text{OC} & 112.0 & 99.9503 & Ref. \protect\cite{js04}\\
\ce{\text{CH2}} & \text{OC} & \text{CH2} & 112.0 & 120.0198 & Ref. \protect\cite{js04}\\
\ce{\text{CH2}} & \text{CH2S} & \text{S} & 114.7 & 100.0 & Ref. \protect\cite{nl05}\\
\ce{\text{OH}} & \text{CH2OH} & \text{CH2} & 109.5 & 100.1490 & Ref. \protect\cite{bc01}\\
\ce{\text{CH2OH}} & \text{CH2} & \text{OC} & 112.0 & 99.9503 & Ref. \protect\cite{js04}\\
\ce{\text{OC}} & \text{CH2} & \text{CH2S} & 112.0 & 99.9503 & Ref. \protect\cite{nl05}\\
\hline
\end{tabular}
\end{center}
\end{table}

\begin{table}[h]
\bibpunct{}{}{,}{n}{,}{,}
\centering
\caption{TraPPE torsions for thiolated PEG which have the form $V_\mathrm{torsion}(\phi) = c_{0} + c_{1}(1+\cos\phi) + c_{2}(1-\cos2\phi) + c_{3}(1+\cos3\phi)$.}
\label{tab:torsion}
\begin{tabular}{ cccc|ccccl }
\toprule
$i$&$j$&$k$&$l$ & $c_0$ & $c_1$ & $c_2$ & $c_3$ & Source\\
\midrule
\ce{\text{CH2}} & \text{CH2OH} & \text{OH} &\text{HO} & 0.0 & 0.4169 & -0.0580 & 0.3734 & Ref. \protect\cite{bc01}\\
\ce{\text{CH2}} & \text{CH2} & \text{OC} &\text{CH2} & 0.0 & 1.4413 & -0.3254 & 1.1092 & Ref. \protect\cite{js04}\\
\ce{\text{OC}} & \text{CH2} & \text{CH2} &\text{OC} & 1.0 & 0.0 & -0.500  & 2.0 & Ref. \protect\cite{js04}\\
\ce{\text{S}} & \text{CH2S} & \text{CH2} &\text{OC} & 1.0 & 0.0 & -0.500  & 2.0 & Ref. \protect\cite{nl05}\\
\ce{\text{OH}} & \text{CH2OH} & \text{CH2} &\text{OC} & 1.0 & 0.0 & -0.500 & 2.0 & Ref. \protect\cite{js04}\\
\ce{\text{CH2OH}} & \text{CH2} & \text{OC} &\text{CH2} & 0.0 & 1.441 & -0.325 & 1.109 & Ref. \protect\cite{nl05}\\
\ce{\text{CH2}} & \text{OC} & \text{CH2} &\text{CH2S} & 0.0 & 1.441 & -0.325 & 1.109 & Ref. \protect\cite{nl05}\\
 \bottomrule
\end{tabular}
\end{table}

\begin{sidewaystable}[h]
\bibpunct{}{}{,}{n}{,}{,}
\centering
\caption{Gold EAM parameters for $\phi(r)$ and $f(r)$. $\phi(r)$ is a generalized elemental pair potential. The electron density function $f(r)$ has the same form as the attractive portion of $\phi(r)$.  
\newline $\phi(r) = \frac{\text{A}\text{exp}[-\alpha(\frac{r}{r_e}-1)]}{1+(\frac{r}{r_e}-\kappa)} - \frac{\text{B}\text{exp}[-\beta(\frac{r}{r_e}-1)]}{1+(\frac{r}{r_e}-\lambda)^{20}}$;
\newline $f(r)=\frac{{f_e}\text{exp}[-\beta(\frac{r}{r_e}-1)]}{1+(\frac{r}{r_e}-\lambda)^{20}}$.} 
\label{tab:eam_phi}
\begin{tabular}{ |c c c c c c c c c c c| }
 \toprule
$r_{e}$(\AA{}) & $f_{e}$(eV/\AA{}) & $\rho_{e}$(eV/\AA{}) & $\rho_{s}$(eV/\AA{}) & $\alpha$ & $\beta$ & A & B & $\kappa$ & $\lambda$ & Source\\
\hline
\ce2.885034 & 1.529021 & 19.991632 & 19.991509 & 9.516052 & 5.075228 & 0.229762 & 0.356666 & 0.35657 & 0.748798 & Ref. \protect\cite{xwz04}\\
\hline
\end{tabular}
\end{sidewaystable}

\begin{sidewaystable}[h]
\bibpunct{}{}{,}{n}{,}{,}
\centering
\caption{Gold EAM parameters for the embedding function $F(\rho)$. The three equations below allow the calculation of $F(\rho)$ at three different electron density ranges:
\newline $F(\rho)=\sum_{i=0}^{3}F_{ni}(\frac{\rho}{\rho_n}-1)^i$, $\rho < \rho_{n}$, $\rho_{n}=0.85\rho_e$;
\newline $F(\rho)=\sum_{i=0}^{3}F_{i}(\frac{\rho}{\rho_e}-1)^i$, $\rho_{n} \leq \rho < \rho_{0}$, $\rho_{0}=1.15\rho_e$;
\newline $F(\rho)=F_{e}[1-ln(\frac{\rho}{\rho_{s}})^{\eta}](\frac{\rho}{\rho_{s}})^{\eta}$, $\rho < \rho_{n}$, $\rho_{0} \leq \rho$.} 
\label{tab:eam_rho}
\begin{tabular}{ |c c c c c c c c c c c| }
 \toprule
$F_{n0}$ & $F_{n1}$ & $F_{n2}$ & $F_{n3}$ & $F_{0}$ & $F_{1}$ & $F_{2}$ & $F_{3}$ & $F_{e}$ & $\eta$ & Source\\
\hline
\ce-2.937772 & -0.500288 & 1.601954 & -0.835530 & -2.98 & 0 & 1.706587 & -1.134778 & -2.978815 & 1.021095 & Ref. \protect\cite{xwz04}\\
\hline
\end{tabular}
\end{sidewaystable}

\clearpage
\begin{table}[h]            
\bibpunct{}{}{,}{n}{,}{,}
\begin{center}
\caption{Gold DR-EAM self-potential parameters. The chargeMass term is a tunable parameter that determines the speed at which the polarizable gold atom responds to an external field. $N_\text{eff}$ describes the effective number of charge carriers per gold atom. All parameter units result in a self-potential in eV. The self-potential is described by $V_{self}(q)= \sum_{n=1}^6 a_{n}q^{n}$.}   
\label{tab:dream}
\begin{tabular}{ |c c c c c c c c c| }     
\hline                 
chargeMass & $N_\mathrm{eff}$ & $a_1$ & $a_2$ & $a_3$ & $a_4$ & $a_5$ & $a_6$ &Source\\
\hline
\ce1000.0 & 0.59 & 13.89 & 7.92 & 13.192055 & 68.173897 & -88.029873 & 28.855093 & Ref. \protect\cite{hb19}\\
\hline
\end{tabular}
\end{center}
\end{table}

\subsection{System Preparation and Equilibration}

Planar gold slabs ($a = 4.08 \mathrm{\AA}$) exposing the (111), (110), and (100) facets were built with slabBuilder, an OpenMD utility that rotates the specified facet so that it is positioned normal to the $z$-axis of the simulation cell.\cite{openmd} An identically-sized solvent box was constructed using Packmol\cite{pack} and OpenMD,\cite{openmd} and the thiolated PEG positions were randomized in each statistically-independent replica (with four replicas for each studied interface). 

The spherical gold nanoparticles used in all non-periodic simulations were built with a lattice constant of 4.08 \AA{} using nanoparticleBuilder, another OpenMD utility.\cite{openmd}  These particles were constructed with a radius of either 10 or 20 \AA{}. The gold nanoparticle underwent structural relaxation with incremental temperature increases up to 300~K. The solvent spheres were created with empty center regions to allow for simplified merging with the gold nanoparticles. One gold atom at the center of the spherical nanoparticle was given a large mass ($m = 1.9697 \times 10^6$ amu) to prevent nanoparticle drifting.

For all systems, the metal and solvent layers were combined using other OpenMD utilities, and steepest descent optimization was used to structurally relax each system. Particle velocities were resampled from a Maxwell-Boltzmann distribution at 300~K prior to equilibration.\cite{openmd} The structural makeup of each system studied is shown in Table \ref{tab:makeup}.

  \begin{table}[h]
    \centering
    \begin{tabular}{l | c c | c c c| }
    
     & \multicolumn{2}{c|}{Spheres} & \multicolumn{3}{c|}{Slabs} \\ \cmidrule{2-3} \cmidrule{4-6}
      & 10 \AA{} & 20 \AA{} & (111) & (110) & (100) \\ \midrule
    Metallic structure (in \AA{}) & $r = 10$ & $r = 20$ & $31 \times 36 \times 28$  & $30 \times 35 \times 29$ & $30 \times 36 \times 29$ \\
    Total System size (in \AA{})  & $r \sim 40$ & $r \sim 50$  & $31 \times 36 \times 117$  & $30 \times 35 \times 120$ & $30 \times 36 \times 118$ \\
    Gold volume (\AA{}$^3$) & 4,189 & 33,510 & 29,340 & 28,525 & 28,525\\
    Solvent volume (\AA{}$^3$) & 175,406 & 490,088 & 101,250 & 97,900 & 101,030\\
    $N_\mathrm{Gold}$ & 249 & 1985 & 1728 & 1680 & 1680\\
    $N_\mathrm{Water}$ & 6018 & 17,024 & 2854 & 2790 & 2790\\
    $N_\mathrm{thiolated PEG}$ & 59 & 234 & 104 & 98 & 100\\
    $\rho_{\mathrm{thiolated PEG}}$ ($\text{nm}^{-2}$) & 4.70 & 4.66 & 4.81 & 4.67 & 4.63\\
% $\rho_{\mathrm{water}}$ (g/cm$^3$) & 1.0 & 1.0 & 1.1 & 1.1 & 1.1 \\
\bottomrule
    \end{tabular}
    \caption{Structures simulated in this work}
    \label{tab:makeup}
\end{table}

All systems using the DR-EAM metal potential were simulated directly prior to equilibration using a time step of 0.2 fs, as the charge degrees of freedom in DR-EAM must be simultaneously propagated. A time step of 1 fs was used for all equilibration and flux application steps. The planar systems were first structurally relaxed (NVT), followed by a pressure correction (NPT) with separate barostats for each dimension. This was followed by thermal relaxation (NVT) and equilibration using a microcanonical integrator (NVE). The planar systems were simulated using periodic boundary conditions. The spherical systems were equilibrated for 250 ps using the non-periodic Langevin Hull integrator\cite{cfv11} and velocity resampling at 300~K every ps, and then an additional 200 ps with no resampling.

After equilibration, a thermal flux was applied to each system using the velocity shearing and scaling reverse non-equilibrium molecular dynamics (VSS-RNEMD) methods developed by Kuang and Gezelter.\cite{ks13} A flux was applied until thermal gradients and temperature discontinuities developed. Kinetic energy was exchanged between either two slabs separated in $z$ (planar systems) or two concentric spherical shells (spherical systems). The thermal flux was applied for 5 ns in all systems. For planar systems, the thermal data shown in this work was averaged across the entire 5 ns of flux application. For the nanospheres, the thermal data was averaged across the last 1 ns of flux application. Coupling to the thermal bath was removed in spherical systems so as to prevent interference with the applied flux. In polarizable systems, the fluctuating charges were propagated using a microcanonical (NVE) integrator to mitigate unphysical losses of kinetic energy into the electronic degrees of freedom. 

\begin{figure}[h]
    \centering
    \includegraphics[width=\linewidth]{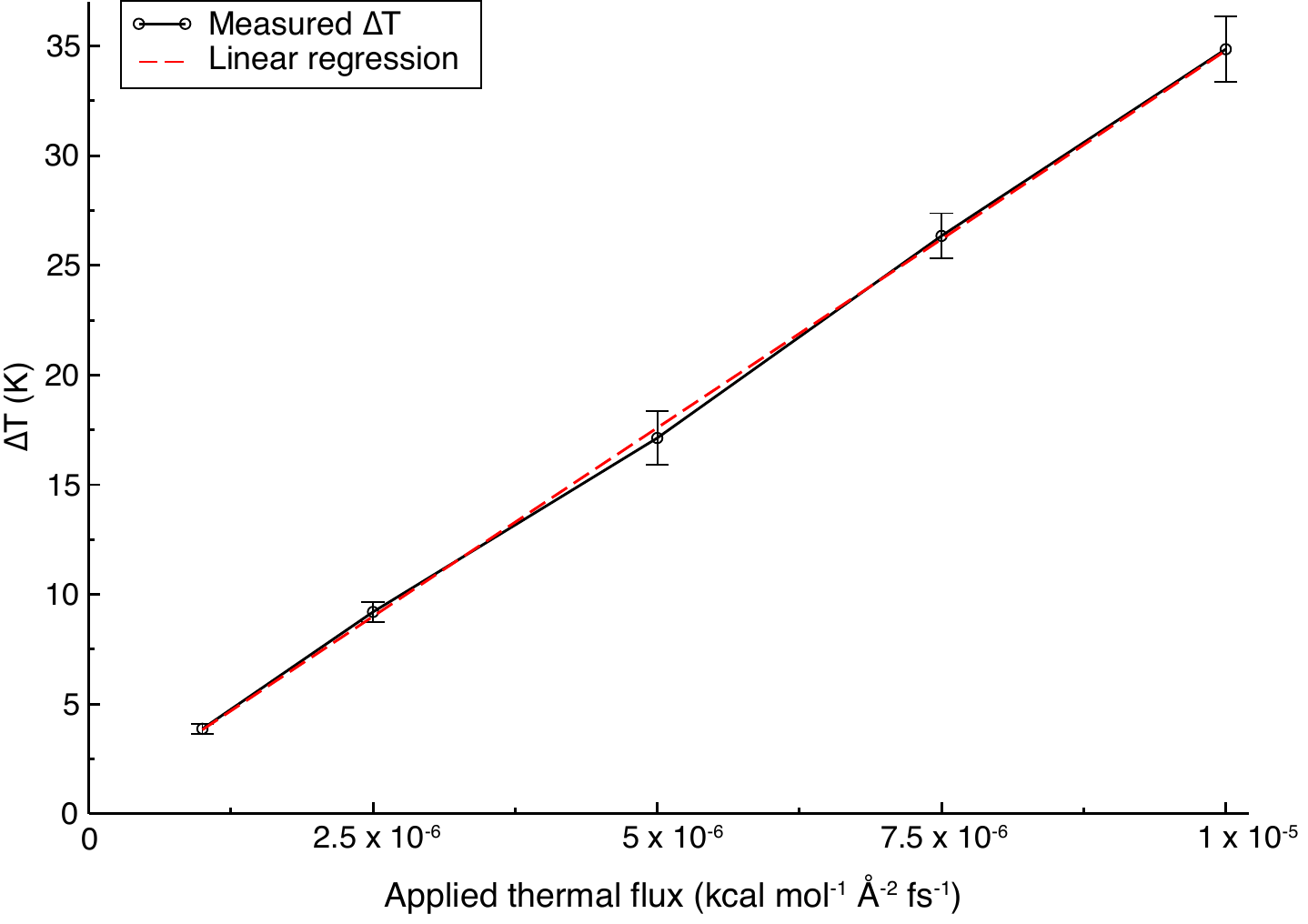}
    \caption{ A test of five values of applied flux bracketing the flux used in the main paper.  This test is for Au(111) interfaces. Each point is an average of four statistically independent simulations.
    \label{fig:fluxdata}} 
\end{figure}

 In Figure \ref{fig:fluxdata}, we show that we are in the linear response regime with the applied flux utilized in the main paper. We tested a total of five flux values, and these were applied for 5 ns on a set of Au(111) systems. The flux used for planar systems in this work was $5.0 \times 10^{-6}$ $\mathrm{~kcal~mol^{-1} \AA{}^{-2}~fs^{-1}}$ $(3500 \mathrm{~MW~m}^{-2})$. 

\subsection{Additional Thermal Profiles and Mass Densities}
\begin{figure}[h]
    \centering
    \includegraphics[width=\linewidth]{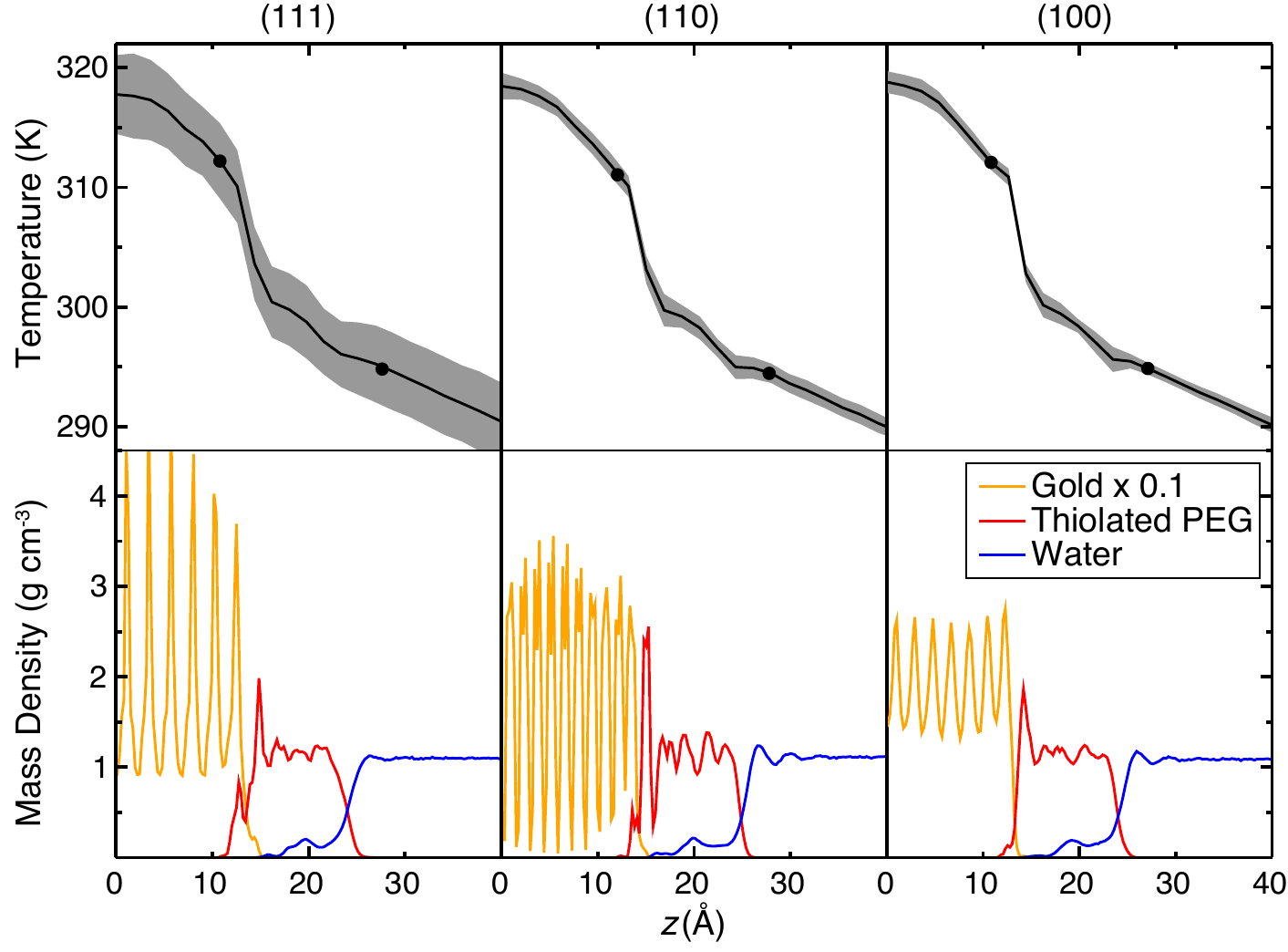}
    \caption{ Under an applied thermal flux, temperature drops are largest at the location of the Gold-Sulfur bond.   (Top) Thermal profiles of planar systems simulated using the polarizable DR-EAM potential. Shading indicates the 95\% confidence interval for computed temperatures, while the black dots are the bounds of the interfacial region used for computing conductance.  (Bottom) Mass densities of each species. $z = 0$ represents the center of the gold slabs.\label{fig:massdensityzfq}}
\end{figure}

\begin{figure}[h]
    \centering
    \includegraphics[width=\linewidth]{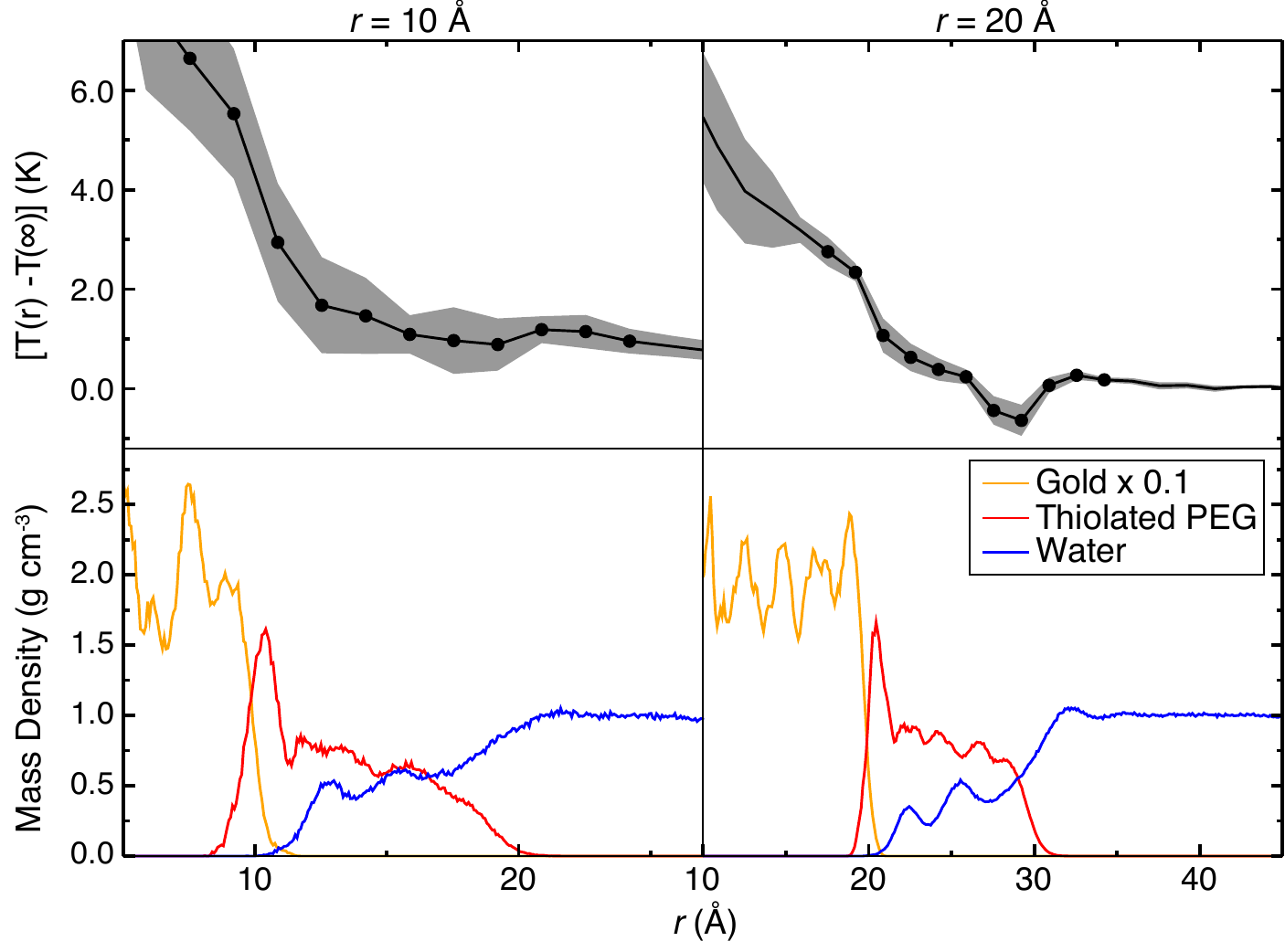}
    \caption{ In the nanospheres, temperature drops are also largest at the Gold-Sulfur bond. (Top) Thermal profiles of nanoparticle systems, relative to the temperature of the solvent far from the particle. The shaded region represents the 95\% confidence intervals. The black data points are the temperature differences in the Kapitza region which were used for calculating conductance.
    (Bottom) Local mass densities of each species for the nanoparticles using the polarizable DR-EAM model. $r = 0$ is at the center of the gold nanosphere. 
    \label{fig:massdensityrfq}} 
\end{figure}

The thermal profiles and local mass densities for all components in polarizable systems are shown in Figures \ref{fig:massdensityzfq} and \ref{fig:massdensityrfq}. The temperature drop is largest between gold and sulfur in these systems as in their non-polarizable counterparts. There are no compelling differences between Figure \ref{fig:massdensityzfq} and Figure \ref{fig:massdensityz} in the main text, but we note a small decrease in the penetration of sulfur into gold in the polarizable Au(111) surface. The ordering of water molecules near the interface are similar to that of Figure \ref{fig:massdensityz}. When comparing Figure \ref{fig:massdensityrfq} with Figure \ref{fig:massdensityr} in the main text, it is clear that polarizability makes no significant difference in the ordering of any species.

\clearpage

\subsection{Additional Vibrational Power Spectra}

\begin{figure}[h!]
    \centering
    \includegraphics[width=\linewidth]{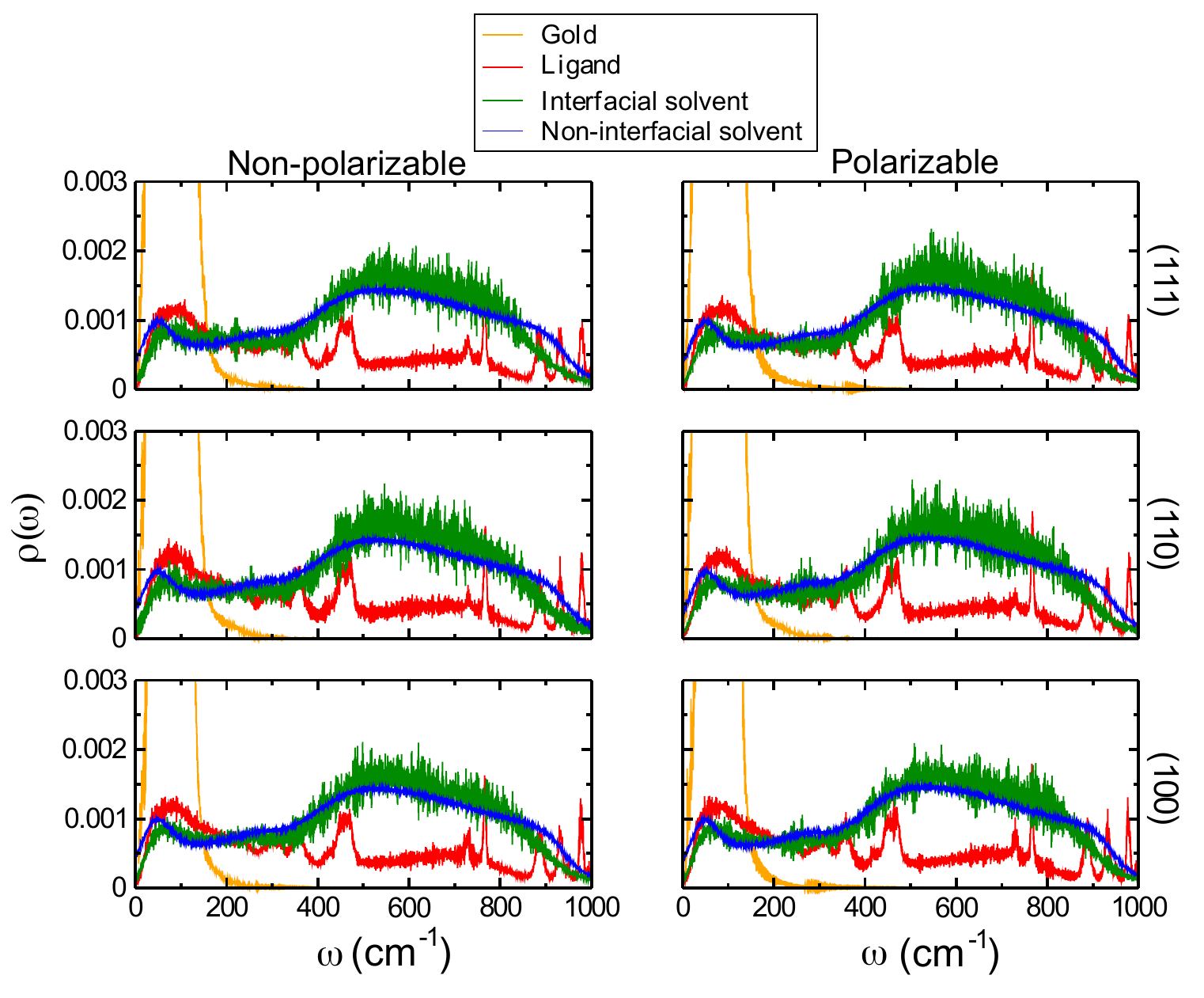}
    \caption{Low frequency portions of the vibrational power spectra for all planar systems studied. Note that the y-axis has been scaled down to allow visualization of the ligand and solvent spectra, and the gold spectra extend beyond the scope of these plots. \label{fig:planarpspect}} 
\end{figure}

The vibrational power spectra for all components of the planar systems are shown in Figure \ref{fig:planarpspect}. There are no clear differences across metallic facet or polarizability here. The full gold spectra cannot be seen here, but they do not differ significantly and obscure the other spectra when in full view.

\begin{figure}[h!]
    \centering
    \includegraphics[width=\linewidth]{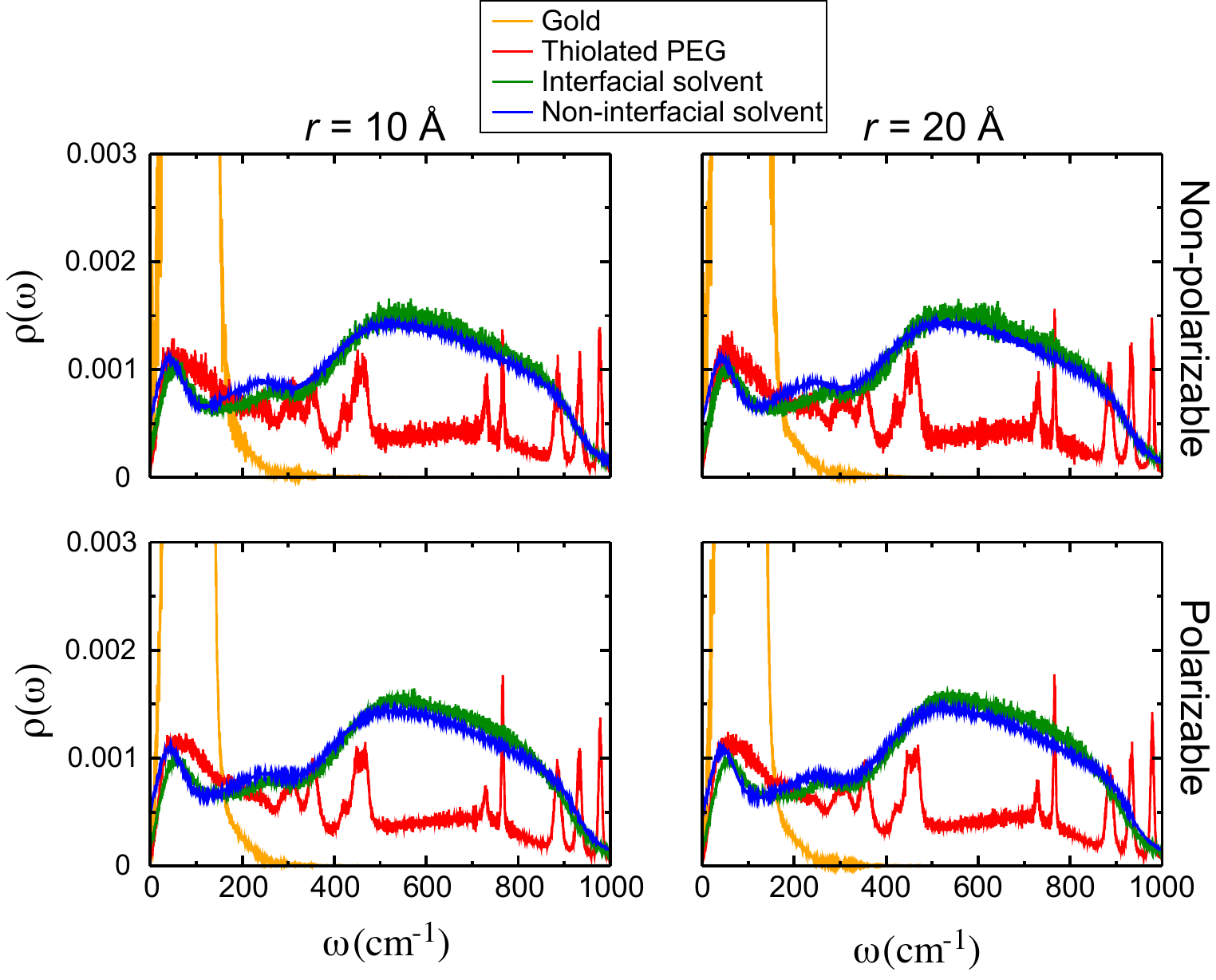}
    \caption{Vibrational power spectra for all spherical systems studied. The y-axis has been scaled as in Fig. \ref{fig:planarpspect}. \label{fig:nppspect}} 
\end{figure}

The vibrational power spectra for all components of spherical systems are shown in Figure \ref{fig:nppspect}. There appears to be little effect of either system size or metal polarizability here as well. Again, the gold spectra do not differ significantly and are thus not fully shown. A full analysis of the differences in spectra between the planar and spherical systems is available in the main text.

\begin{figure}[h!]
    \centering
    \includegraphics[width=\linewidth]{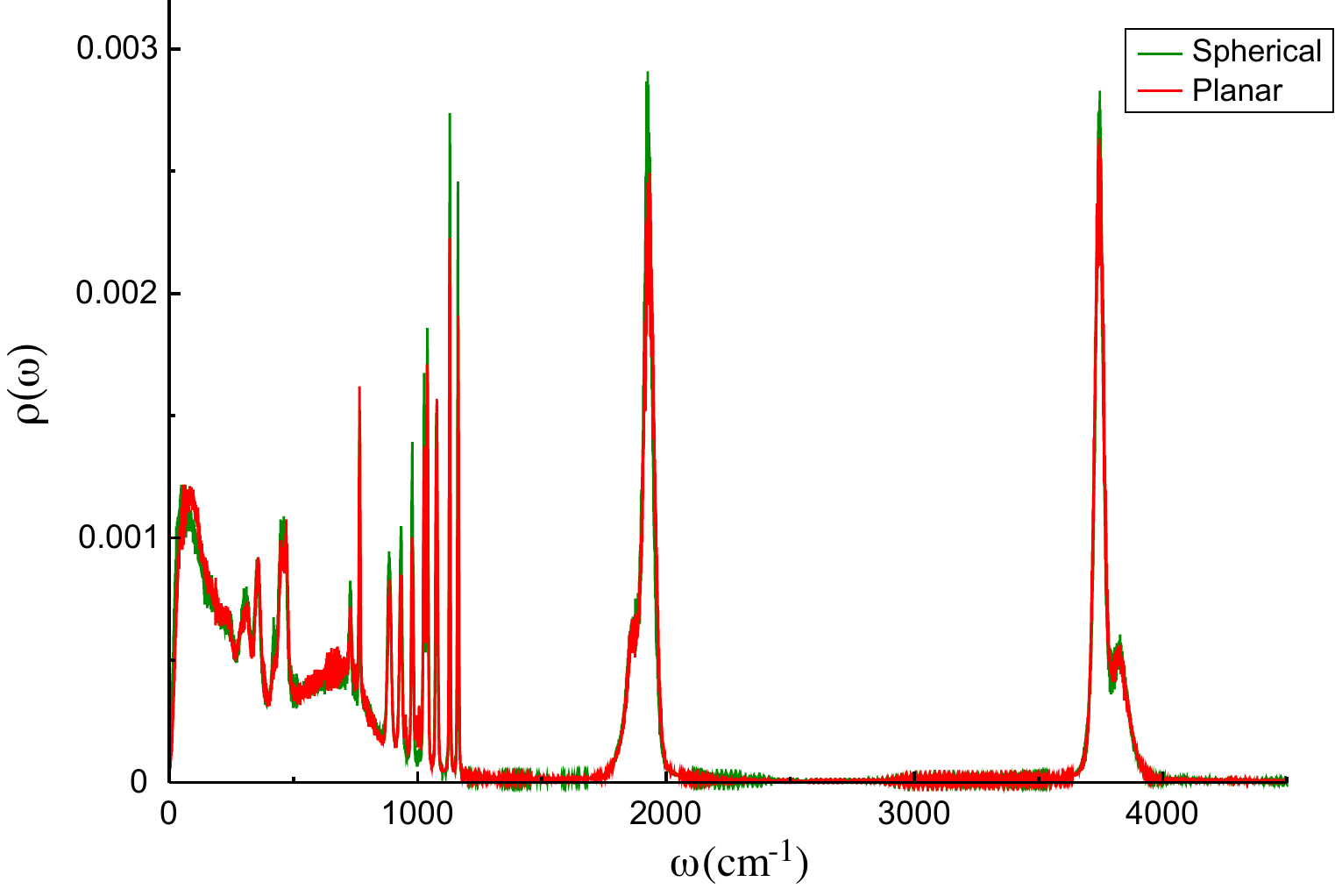}
    \caption{Averaged thiolated PEG vibrational power spectra for planar and spherical systems. Note: the planar curve is an average of all polarizable and non-polarizable (111), (110), and (100) facets, and the spherical curve is an average of all polarizable and non-polarizable nanospheres with radii of 10 and 20 \AA{}. \label{fig:thiolpegfullpspect}} 
\end{figure}

As thiolated PEG exhibits both low and high frequency activity, we have computed the averaged spectra for thiolated PEG in the planar and nanosphere systems which are are shown in Figure \ref{fig:thiolpegfullpspect}. Overall, these spectra do not differ significantly. The peak ranging from approximately 3700 - 3900 $\text{cm}^{-1}$ is likely an O-H stretch from the terminal hydroxyl group.

\clearpage

\subsection{Additional Bhattacharyya Coefficients}

We have also included Bhattacharyya coefficients (BC) which are similar calculations to those in Table \ref{tab:bcm} but using power spectra rather than mass densitites. The BC values for the thiolated PEG / gold and thiolated PEG / water pairs are shown in Table \ref{tab:bcp}. The water has been categorized into interfacial and bulk regions, where interfacial water comprises solvent molecules that are trapped in the thiolated PEG layer, and bulk water captures the solvent molecules outside that region. The magnitude of BC values are larger for each pair here than those for the same pairs calculated with mass densities. These coefficients appear to be unaffected by metallic polarizability or planar facet, but the mean BC values for the thiolated PEG / gold pair are higher in the 10 \AA{} nanospheres than for the other systems. As expected, there is very little difference between the BCs for bulk and interfacial water, and their power spectra exhibit a significant amount of overlap with that of thiolated PEG. In Table \ref{tab:bcm}, we noted that the BCs for the thiolated PEG-gold pair in the Au(111) systems were about twice as large as those for the Au(110) and Au(100) systems, but that difference does not exist here.

\begin{table}[h!]
    \centering
    \begin{tabular}{l | l | c  c  c}
    \multirow{2}{*}{Facet} & \multirow{2}{*}{Metal} &  \multicolumn{3}{c}{$\text{BC}$ with thiolated PEG (unitless)} \\ \cline{3-5}
    & & Interfacial water & Bulk water & Gold\\ \toprule
    \multirow{2}{1.3in}{(111)} & non-polarizable & 0.740 $\pm$ 0.007 & 0.737 $\pm$ 0.007 & 0.434 $\pm$ 0.007 \\
     & polarizable & 0.732 $\pm$ 0.008 & 0.741 $\pm$ 0.014 & 0.447 $\pm$ 0.013 \\ \midrule
    \multirow{2}{1.3in}{(110)}& non-polarizable & 0.737 $\pm$ 0.011 & 0.745 $\pm$ 0.007 & 0.438 $\pm$ 0.018 \\
     & polarizable & 0.735 $\pm$ 0.009 & 0.735 $\pm$ 0.002 & 0.426 $\pm$ 0.006\\ \midrule
     \multirow{2}{1.3in}{(100)} & non-polarizable & 0.747 $\pm$ 0.007 & 0.747 $\pm$ 0.004 & 0.424 $\pm$ 0.013\\
    & polarizable & 0.736 $\pm$ 0.007 & 0.739 $\pm$ 0.007 & 0.423 $\pm$ 0.015 \\ \midrule
    \multirow{2}{1.3in}{NP ($r = 10$~\AA{})} & non-polarizable & 0.739 $\pm$ 0.015 & 0.740 $\pm$ 0.014 & 0.452 $\pm$ 0.022\\
    & polarizable & 0.740 $\pm$ 0.011 & 0.740 $\pm$ 0.013 & 0.452 $\pm$ 0.016 \\ \midrule
    \multirow{2}{1.3in}{NP ($r = 20$~\AA{})} & non-polarizable & 0.745 $\pm$ 0.004 &  0.745 $\pm$ 0.004 & 0.429 $\pm$ 0.008\\
    & polarizable & 0.741 $\pm$ 0.005 & 0.742 $\pm$ 0.004 & 0.427 $\pm$ 0.006\\ \bottomrule
    \end{tabular}
    \caption{Vibrational power spectral overlap measured using Bhattacharyya coefficients, showing the degree of similarity between the thiolated PEG and the other two components of the metal / ligand / water systems.}
    \label{tab:bcp}
\end{table}
\clearpage
\subsection{Charge Penetration at the Interface}

The image charge effect describes the phenomenon in which a metallic species responds to a nearby external charge by inducing the opposite charge on its surface. DR-EAM has previously been shown to replicate this effect in systems of bare gold solvated in water.\cite{hb19, ss22} To investigate how these systems respond when the gold is polarizable, we calculated the charge density as a function of $z$ (for planar systems) or $r$ (for spherical systems). 

\begin{figure}[h]
    \centering
    \includegraphics[width=\linewidth]{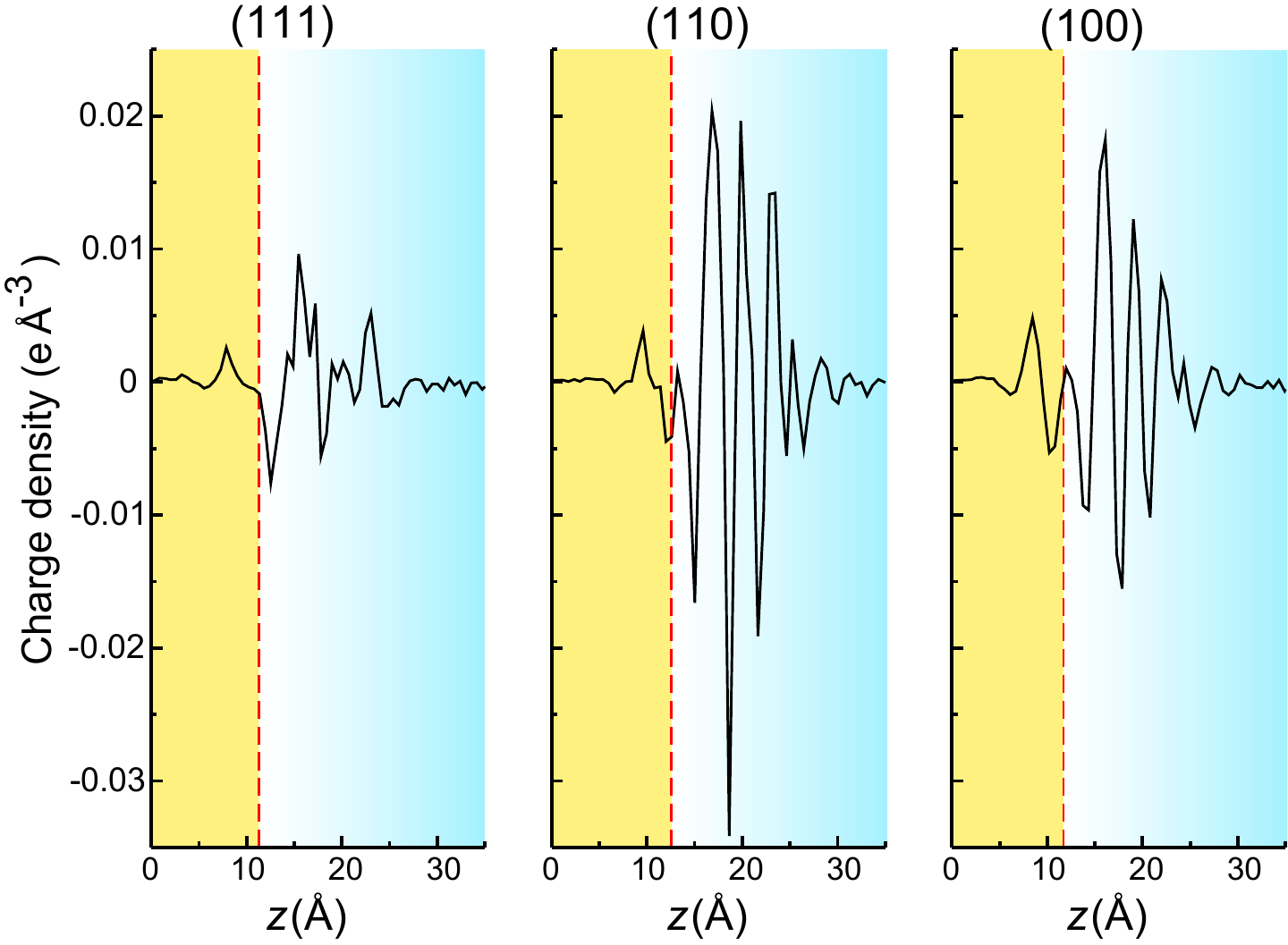}
    \caption{Charge density as a function of $z$ in planar systems using the polarizable metal potential DR-EAM. The dashed red line indicates the gold/thiolated PEG interface in each plot. \label{fig:chargez}} 
\end{figure}

Figure \ref{fig:chargez} shows the charge density as a function of $z$ for planar systems. The positive peak directly to the left of the interface on each plot (dashed red line) represents the positive charge of the surface gold atoms responding to the negatively charged sulfur atoms near the surface. The magnitude of these peaks does not vary significantly, suggesting that the difference in gold atom packing across fcc lattices does not change the ordering of charges at the gold surface. The negative peak directly adjacent to the gold represents the sulfur atoms in the thiolated PEG molecules. Unlike the gold peaks, these vary across gold facets, as does the magnitude of the ether oxygen peaks starting at around 20 \AA{}. This difference in the gold surface disruption likely leads to the differing magnitudes of sulfur and ether oxygen peaks. It is also clear from these plots that the ligand layer is more ordered on the Au(110) and Au(100) facets. 

\begin{figure}[h]
    \centering
    \includegraphics[width=\linewidth]{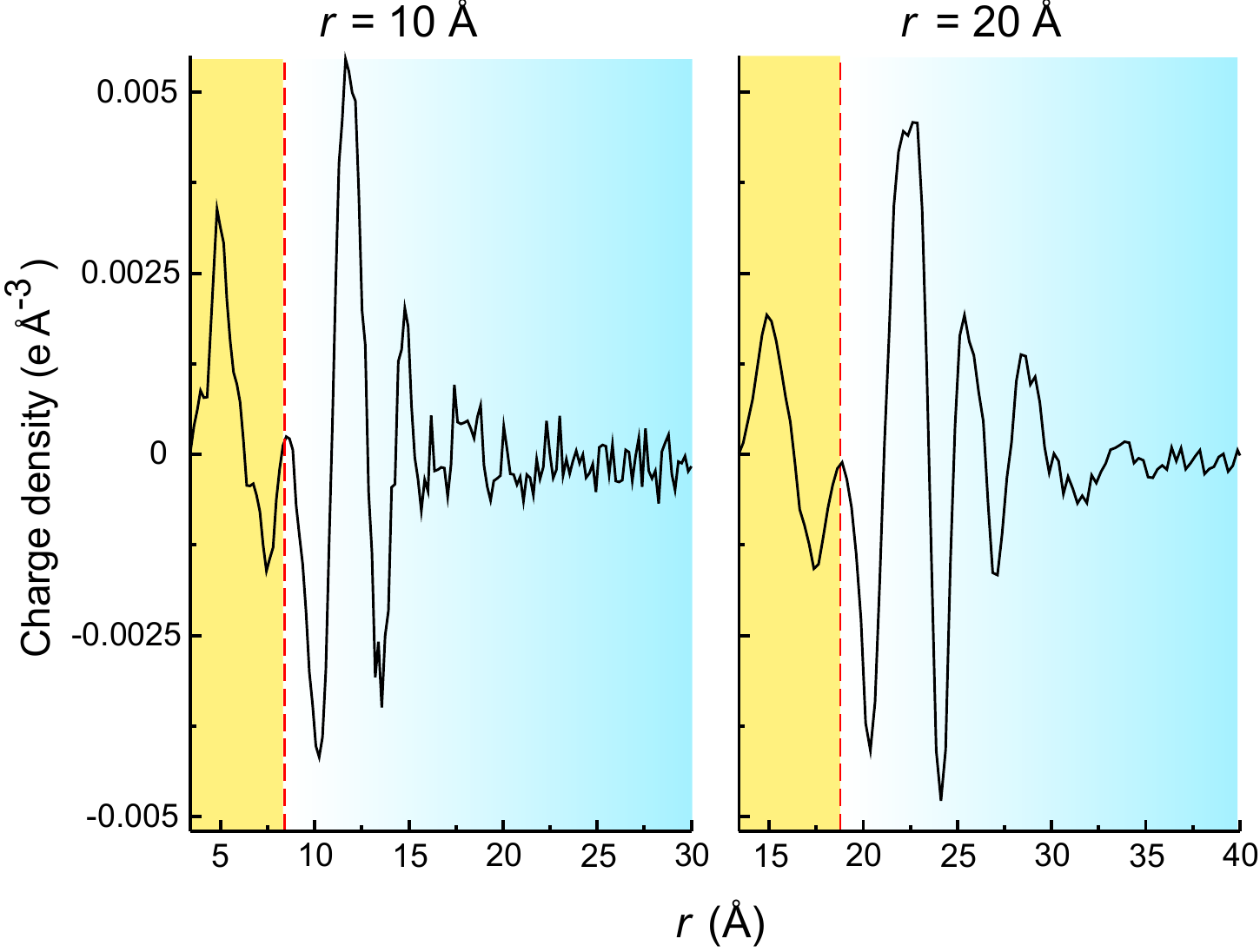}
    \caption{Charge density as a function of $r$ in spherical systems using the polarizable metal potential DR-EAM. \label{fig:charger}} 
\end{figure}

The charge density as a function of $r$ is shown for spherical systems in Figure \ref{fig:charger}. Here, the magnitude of the first gold peak is approximately equal in both systems, as is the sulfur peak. However, the magnitude of the second gold peak differs greatly, with a much larger magnitude in the smaller systems. Additionally, the ether oxygens appear to be more ordered in systems where the gold radius is 20 \AA{}, with clear ordering out to $r = 30$ \AA{}, the approximate location of the hydroxyl oxygen. In the smaller nanospheres, the thiolated PEG chains have more internal disorder, and the ether oxygen minima are discernible only out to $r = 15$ \AA{}. 

\clearpage
\newpage
\bibliography{ThiolPEG}